\documentclass[% Try both options
  aps,prr,twocolumn,showpacs,amsmath,amssymb,superscriptaddress
]{revtex4-1}

\makeatletter
\newcommand{\whencolumns}[2]{\preprintsty@sw{#1}{#2}}
\makeatother

\usepackage[english]{babel}
\usepackage{amsmath,amssymb,amsfonts} 	% Typical maths resource package
\usepackage{graphicx}
\usepackage{array,multirow}
\usepackage[utf8]{inputenc}
\usepackage{bm}
\usepackage{xcolor}
\usepackage[normalem]{ulem}
\usepackage{siunitx}
\usepackage{hyperref}
\hypersetup{colorlinks=true,linkcolor=blue,citecolor=blue,urlcolor=blue}
\usepackage{url}
\usepackage{textcomp}
\usepackage{lineno,hyperref}
\usepackage{float}
\usepackage{textcomp} %for n^o
\usepackage{rotating}
\usepackage[figurename=FIG.]{caption}
\usepackage{subcaption}
\usepackage{color} % only to put comments in the draft
\definecolor{red}{rgb}{1.,0.,0.}
\usepackage{soul}
\usepackage{amsmath}
\usepackage{enumitem}

 \usepackage[
 justification=raggedright]
 {caption}
\usepackage{subcaption}
\usepackage{qcircuit}
\usepackage{braket}

\makeatletter
\newcommand*{\rom}[1]{\expandafter\@slowromancap\romannumeral #1@}
\makeatother

\newcommand\THEOS{Theory and Simulation of Materials (THEOS), \'Ecole Polytechnique F{\'e}d{\'e}rale de Lausanne, 1015 Lausanne, Switzerland}
\newcommand\MARVEL{National Centre for Computational Design and Discovery of Novel Materials (MARVEL), {\'E}cole Polytechnique F{\'e}d{\'e}rale de Lausanne, 1015 Lausanne, Switzerland}

\newcommand\IBM{IBM Quantum, IBM Research—Zurich, R\"uschlikon 8803 Switzerland}

\begin{document}
%To be discussed
\title{Effective calculation of the Green's function in the time domain on near-term quantum processors}
%Novel quantum algorithm for the effective calculation of the green's funciton in the time domain

\author{Francesco Libbi}
\email[Corresponding author. ]{francesco.libbi@epfl.ch}
\author{Jacopo Rizzo}
\affiliation{\THEOS}\affiliation{\MARVEL}\affiliation{\IBM}
\author{Francesco Tacchino}
\affiliation{\IBM}
\author{Nicola Marzari}
\affiliation{\THEOS}\affiliation{\MARVEL}
\author{Ivano Tavernelli}
\affiliation{\IBM}
%Referee
%Calandra
%Davide prendergast
%Roxana margine
%Gianmarco Rignanese

\begin{abstract}
We propose an improved quantum algorithm to calculate the Green's function through real-time propagation, and use it to compute the retarded Green's function for the 2-, 3- and 4-site Hubbard models. This novel protocol significantly reduces the number of controlled operations when compared to those previously suggested in literature. Such reduction is quite remarkable when considering the 2-site Hubbard model, for which we show that it is possible to obtain the exact time propagation of the $\ket{N\pm 1}$ states by exponentiating one single Pauli component of the Hamiltonian, allowing us to perform the calculations on an actual superconducting quantum processor. 

\end{abstract}

\maketitle
%\linenumbers

\section{introduction}

The many-body Green's function of an interacting quantum system is a central quantity in many  state-of-the-art techniques for solid state physics and quantum chemistry. 
For instance, the GW method \cite{10.3389/fchem.2019.00377,Aryasetiawan_1998} and higher order diagrammatic expansions of the self-energy \cite{PhysRevB.92.081104} use the Green's function to correct the DFT band structure of solids, while dynamical mean-field theory (DMFT) \cite{RevModPhys.68.13} tackles strongly correlated materials. Furthermore, non-equilibrium Green's functions \cite{Stefanucci} are used to study the ultra-fast dynamics of systems which are excited e.g. by a strong pulse.

The exact many-body Green's function possesses many important properties. 
First, it can easily be connected to observables (e.g. the density of the system $n(x,t)$ is equal to its diagonal component), 
and the poles of its Fourier transform correspond to the excitation energies. Finally, the retarded Green' s function is connected to the exact linear response of the system. 

The computation of the exact Green's function on a classical computer is known to become exponentially hard when increasing the size of the system under study, as it requires the knowledge of the exact ground state. It becomes thus attractive to devise algorithms for the calculation of Green's functions on quantum computers, since even though quantum algorithms make use of the many-electron wavefunction to evaluate the Green's functions, there is no actual need of measuring its components. At  variance with equivalent classical approaches based on the Lehman representation, the wavefunction is first mapped to the quantum circuit, then optimized and finally used -- without explicitly storing it full classical description obtained, for instance, through state tomography -- to probe the Green's function. Several works presented in literature are based on the quantum phase estimation \cite{PhysRevA.103.032404,PhysRevA.92.062318,PhysRevC.100.034610,PhysRevA.101.012330} or on the Suzuki-Trotter expansion of the propagator \cite{PhysRevX.6.031045,Kreula2016,Chiesa2019}. However, both of these techniques are known to be expensive in terms of circuit depth and number of controlled operations, and therefore are not suitable for near-term quantum computers. One of the first efforts to design an affordable quantum algorithm for the calculation of the Green's function goes back to Endo et al.~\cite{PhysRevResearch.2.033281}. In their work these authors suggest two alternative techniques, one in the frequency domain and the other in the time domain. The former exploits the Lehman representation and calculates the excitation energies of the system through the subspace-search variational quantum eigensolver (SSVQE) \cite{PhysRevResearch.1.033062}, and an alternative approach has been recently presented in Ref.~\cite{rizzo2022oneparticle}, where the SSVQE is replaced with a generalized quantum equation of motion (qEOM) technique \cite{PhysRevResearch.2.043140} for charged excitations.
The second approach of  Ref.~\cite{PhysRevResearch.2.033281} is based on the real time evolution of a system through the variational quantum simulation (VQS) technique \cite{Yuan2019theoryofvariational}, which requires shallower quantum circuits compared to the Suzuki-Trotter expansion of the propagator. The algorithm of Ref.~\cite{PhysRevResearch.2.033281} requires the adoption of a variational form $U$ which can accurately describe the time evolution of both the ground state of the system $\ket{\psi}$ and a state of the form $\ket{P_i\psi}$ (where $P_i$ is a $n$-qubit Pauli operator) with the same choice of parameters. However, as shown in Sec.~\ref{III}, it is quite difficult to find an empiric variational form satisfying this requirement. In addition, such variational form is expected to require deeper circuits than the one needed to evolve one single quantum state.\\
In this work, we propose a novel algorithm that make use of a variational form $U$ evolving only the state $\ket{P_i\psi}$, and show that it leads to much shallower circuits.
%As for the other works in literature, this algorithm exploits the ground state of the Hamiltonian which is physically stored in the quantum register, and does not require its tomography. 
The paper is organized as follows. In Sec. \ref{II} we introduce the definition of the components of the Green's function, and show how these can be calculated through real-time evolution techniques. In Sec. \ref{III} we discuss the choice of the ansatz for the VQS algorithm, listing the most important requirements and suggesting possible strategies to comply with these. Eventually we will see how the choice of a variational form falls naturally upon the variational Hamiltonian ansatz~\cite{PhysRevA.92.042303,2019}. In Sec.~\ref{IV} we present our novel algorithm for the calculation of the Green's function, and in Sec. \ref{V} we compare the results obtained from numerical simulations with those from the algorithm of Ref.~\cite{PhysRevResearch.2.033281} for the case study of a 2-site Hubbard model. We show that for this particular problem the exact evolution of the state $P_i\ket{\psi}$ can be obtained by exponentiating only \textit{one} of the Pauli components of the qubit Hamiltonian, while the same simplification does not hold when evolving $\ket{\psi}$ and $\ket{P_i\psi}$ at the same time. The quantum circuit corresponding to the algorithm suggested is shallow enough to be executed on current quantum computers. In Sec.~\ref{VI} we compare the results obtained with the two algorithms for the 3-site Hubbard model with open-boundary conditions and 4-site Hubbard model with periodic-boundary conditions, and discuss the scaling for larger problems. Finally, in Sec.~\ref{VII} we show the results for the 2-site Hubbard model obtained with state-of-the-art superconducting IBM Quantum processors.

 %\textcolor{red}{At this point, I will anticipate (in a new section or as part of the next one) the use of the time-propagation for the calculation of the Green's functions; at least the first part of section III. This will make cleared why are you looking into time propagation.}

\section{Green's function through real-time evolution}\label{II} 
Before presenting the algorithm for the calculation of the correlations, it is useful to define the components of the Green's function of relevance for this work. 
The lesser Green's function is defined as 
\begin{equation}
    G_{lm}^<(t) = -i\braket{ \psi| e^{iHt}c_le^{-iHt} c_m^\dag |\psi }\ ,
\end{equation}
where $\ket{\psi}$ is the ground state of the Hamiltonian and $c_m^\dag, c_l$ are fermionic creation and annihilation operators respectively. The greater component is instead 
\begin{equation}
    G_{lm}^>(t) = +i\braket{ \psi| c_m^\dag e^{iHt}c_le^{-iHt}  |\psi }\ .
\end{equation}
The difference between these components is the retarded Green's function, which is the target of our calculations for its applications in linear-response theory:
\begin{equation}
    G_{lm}^R = 
    \Bigl[ G_{lm}^<(t) - G_{lm}^>(t) \Bigr]\theta(t)\ , 
\end{equation}
where $\theta(t)$ is the Heavyside step function.
The real-time approach to the calculation of the Green's function consists in computing the states $e^{-iHt}\ket{\psi}$ and $e^{-iHt}c^\dag_m\ket{\psi}$, which represent the time evolution of the states $\ket{\psi}$ and $c^\dag_m\ket{\psi}$ respectively. The choice of the time evolution technique, which is crucial for designing an efficient quantum circuit for the calculation of the Green's function, is discussed in the next paragraph.  \\
 
\section{Choice of the variational form for the VQS}\label{III}
We briefly introduce the techniques suitable to follow the time evolution of a quantum system, and then focus on the choice of a proper ansatz for the VQS.
The time evolution of a generic quantum state $\ket{\phi_0}$ is obtained through the action of the time evolution operator:
\begin{equation}
\ket{\phi(t)} = e^{-iHt} \ket{\phi_0}\ .
\end{equation}
On a quantum computer, this can be achieved by applying the unitary gate expressing the exponential of the Hamiltonian to the qubit register encoding the state $\ket{\phi_0}$. The qubit Hamiltonian, which is built thanks to the mapping of the fermionic Hamiltonian to the qubit Hilbert space, is a weighted sum of multiple-qubits Pauli gates
\begin{equation}
    H = \sum_m c_m P_m \ .
\end{equation}
Therefore, the exact expression of the propagator is
\begin{equation}
    U = e^{-i \sum_m c_m P_m t}\ .
\end{equation}
Since the operators $P_m$ does not necessarily commute with each other, it is not possible to express $U$ as a product of exponentials in an exact way. Nevertheless, thanks to the Suzuki-Trotter expansion, the propagator can be approximated as a product of exponentials:
\begin{equation}\label{trotter}
    e^{-i \sum_m c_m P_m t} = \Bigl( \prod_m e^{-i c_m P_m t /n} \Bigr)^n + o\Bigl(\frac{t^2}{n}\Bigr) \ .
\end{equation}
Unfortunately, this approach requires deep circuits, especially when the time evolution is carried for long times, and therefore it is not optimal for near-term noisy quantum devices. A remarkable advantage in term of circuit depth can instead be obtained by adopting variational quantum simulation (VQS) algorithms~\cite{Yuan2019theoryofvariational,2021}. These techniques consist in optimizing the parameters of a variational form $U(\bar{\theta}=\theta_1,...,\theta_M)$ to satisfy the variational principles which hold for the exact time evolution. The  variational principle most often adopted for quantum simulations is McLachlan's, due to its numerical stability, and reads
\begin{equation}\label{mclachlan}
    \delta||d/dt + iH \ket{\phi(\bar{\theta}(t))} || = 0 \ . 
\end{equation}
Eq. \ref{mclachlan} leads to the following differential equation for the parameters $\bar{\theta}$:
\begin{equation}\label{mclachlan1}
    \sum_{j=1} M_{ij}\dot{\theta}_j = V_i\ ,
\end{equation}
where
\begin{equation} \label{M}
    M_{ij} = \Re{\braket{\partial_i\phi|\partial_j\phi}} 
    + \braket{\partial_i \phi|\phi} \braket{\partial_j \phi|\phi}\ ,
\end{equation}
and
\begin{equation} \label{Vi}
    V_{i} = \Im{\braket{\partial_i\phi|H|\phi}} + i \braket{\partial_i\phi|\phi}\braket{\phi|H|\phi} \ .
\end{equation}
Here, we have adopted the shortcut notations $\ket{\phi} = \ket{\phi(\bar{\theta}(t))}$ and $\ket{\partial_i\phi} = \frac{\partial}{\partial \theta_i}\ket{\phi(\bar{\theta}(t))}$.\\
The McLachlan variational principle has a simple and instructive geometric interpretation, which is discussed in Ref.~\cite{10.21468/SciPostPhys.9.4.048}: the variational form $U(\theta_1,...\theta_M)$ represents a M-dimensional manifold embedded in the $N$ dimension Hilbert space ($N$ being equal to $2^n$, with n number of qubits). In most cases the parameters $\theta_i$ are real, and therefore it is more convenient to describe the Hilbert space as a $2N$-dimensional real vector space, i.e. a space which is spanned by $2N$ \textit{complex} vectors with \textit{real} coefficients (see Ref.~\cite{10.21468/SciPostPhys.9.4.048} for a detailed discussion). In a real vector space, the tangent space to the manifold $\mathcal{M}$ associated to the variational form $U(\bar{\theta})$ is spanned by the vectors:
\begin{equation}\label{Ti}
    \ket{T_i} = \ket{\partial_i\phi} - \braket{\phi|\partial_i\phi}\ket{\phi}\ .
\end{equation}
It is easy to check that $\braket{T_i|\phi}= 0,\, \forall i=1,...M$, as $\ket{\phi}$ belongs to the subspace of vectors with norm 1. 
In Appendix A, we show that Eq. \ref{mclachlan1} can be elegantly written as
\begin{equation}\label{geometric}
    \sum_{j=1}^M \Re{\braket{T_i|T_j}} \dot{\theta}_j = \Im{\braket{T_i|H|\phi}} ,
\end{equation}
with $\braket{T_i|T_j}$ being the Quantum Geometric Tensor~\cite{KOLODRUBETZ20171,PhysRevX.9.011034}. Eq. \ref{geometric} states that only the component of the vector $H \ket{\phi}$ which is parallel to the tangent subspace to the manifold at the point $\ket{\phi}$ drives the time evolution.
Note that when $H \ket{\phi}$ is perpendicular to the tangent subspace at the manifold, the right hand side of Eq. \ref{geometric} vanishes, so there is no time evolution (apart form a global phase). This happens correctly when $\ket{\phi}$ is eigenstate of the Hamiltonian, and $\braket{T_i|H|\phi} = E\braket{T_i|\phi}=0$. However, this can happen also due to a poor choice of the variational form, leading to an unphysical evolution where the state remains constant. Unfortunately, the choice of a good variational form is in general quite problematic, as we will show in the following.

A variational form is actually a circuit expressing the unitary operator $U(\bar{\theta})$ which is applied to the state $\ket{\phi_0}$ to be evolved. As such, it must satisfy the following conditions. 
(I) It must be equal to the identity for the initial choice of the parameters $\bar{\theta}_0$. 
It is not straightforward to identify the parameters which turn the variational form into the identity, and 
most often it is easier to design variational forms that are equal to the identity for $\bar{\theta}_0=0$. 
(II) The number of parameters must be much smaller than the size of the Hilbert space, otherwise the advantage of the variational algorithm would be lost. 
(III) It must be such that the the projection of the vector $H \ket{\phi}$ (which is proportional to the exact time derivative of the state $\ket{\phi}$) on the tangent subspace is as large as possible, as underscored by Eq. \ref{geometric}. 
(IV) In order to be realizable on near-term quantum computers, it must contain a limited number of controlled operations, and, more in general, lead to sufficiently shallow circuits. 
We can now distinguish two main kinds of variational forms. Empiric ans\"atze are expressive quantum circuits usually characterized by the alternation of CNOTs which create entaglement and rotations for the parametrization~\cite{PhysRevResearch.2.033125}. 
As opposite to empiric ans\"atze, the physically inspired ans\"atze are based on the physical properties of the system. The most important representative of this category is the variational Hamiltonian ansatz (VHA) \cite{PhysRevA.92.042303,2019}.
This takes inspiration from the Suzuki-Trotter expansion of Eq. \ref{trotter}, and the parametrization is introduced by replacing the product $c_mt$ with a function $\theta_m(t)$:
\begin{equation}\label{VHA}
    U_{VHA} = \prod_{d=1}^{n_d} \Bigl( \prod_{m=1}^M e^{i\theta_m(t)P_m} \Bigr) \ ;
\end{equation}
here, $n_d$ is the depth of the circuit, and corresponds to the number of Trotter steps. We will now highlight the issues that can drive from the choice of an empiric ansatz, and show how those can be circumvented by adopting the VHA.

\begin{figure}[h]
    \centering
    \includegraphics[scale=0.4]{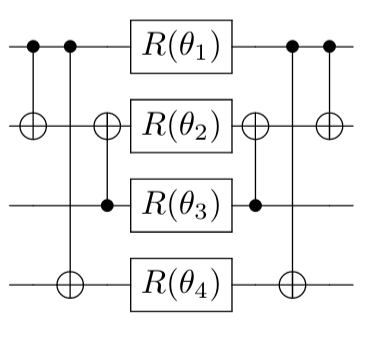}
    \caption{Example of quantum circuit which is equal to the identity when $\bar{\theta}=0$.}
    \label{cnot}
\end{figure}

\textit{The initial value problem}. 
As stated before, any ansatz must contain entangling operators which express an interacting many-body state. One of the most used entangling operator is the CNOT gate, since it is native on many superconducting devices. The position of these CNOTs must be carefully engineered in such a way that the ansatz is equal to the identity at the beginning of the evolution, as required by point I. This is not an easy task to achieve while keeping the circuit shallow at the same time. A strategy to do so can exploit the identity $\mathrm{CNOT^2=I}$ and compose a circuit of units such as those represented in Fig.~\ref{cnot}, where a layer of CNOTs is followed by a layer of rotations which are equal to the identity for $\theta=0$ and then another layer of CNOTs with the order inverted. We will see that this kind of structure looks very similar to the circuit to exponentiate the single Pauli operator $P_m$ of the qubit Hamiltonian.  
Another possibility is to use an ansatz of the following form:
\begin{equation}
    U = \Tilde{U}^\dag(\bar{\theta_0})\Tilde{U}(\bar{\theta})\ ,
\end{equation}
which guarantees by construction condition I. However, both strategies require to double the number of CNOTs, and are thus less suited for proof-of-principle demonstrations on near-term quantum processors (point IV).

\textit{The problem of the initialization of the time evolution}. As anticipated above, the VQS can predict erroneously the initial state to remain constant in the time, if the vector $H\ket{\phi_0}$ is orthogonal to the tangent subspace at the point $\ket{\phi_0}$ of the manifold associated to the variational form. Unfortunately, this can happen often, and represents a major issue in the choice of an empiric ansatz. The reason is simple: the vector $H\ket{\phi_0}$ points to one of the $2N$ possible directions of the real vector space associated to the Hilbert space. Choosing an empiric ansatz of $M$ parameters, with $M$ considerably smaller than $N$ (in order to satisfy point II), consists in drawing a $M$-dimensional surface in a much larger space, while hoping that this surface be parallel to a specific direction, which is likely to fail. Things are made more difficult by the fact that different VQS parameters may span the same direction in the tangent subspace. An example of this is represented by the ans\"atze with tunable depth
\begin{equation}
    U(\bar{\theta}) = \prod_{d=1}^{n_d} \Tilde{U}(\theta_1^d, ..., \theta_l^d)\, ,
\end{equation}
 which are obtained by repeating the same block $\Tilde{U}$ a number of times equal to the circuit depth $n_d$. It can be easily proven that increasing the circuit depth does not increase the size of the tangent subspace: as expressed in Eq. \ref{Ti}, the tangent subspace is generated by the derivatives of the variational form with respect to its parameters; if we consider two corresponding parameters $\theta_i^{d_1}$ and $\theta_i^{d_2}$ belonging to the layers $d_1$ and $d_2$, we can prove that the derivative of the ansatz U with respect to these two parameters are equal to each other at the origin (in the case in which this coincides with the point $\bar{\theta}=0$):
\begin{equation}
\begin{aligned}
    \frac{\partial U}{\partial \theta_i^{d_1}}  (\bar{\theta}=0) = \Tilde{U}(0)\ ...\  \frac{\partial \Tilde{U}}{\partial \theta_i^{d_1}} (0)\ ... \ \Tilde{U}(0) = \frac{\partial \Tilde{U}}{\partial \theta_i^{d_1}} (0)\\ = \frac{\partial \Tilde{U}}{\partial \theta_i^{d_2}} (0) =
    \frac{\partial U}{\partial \theta_i^{d_2}} (\bar{\theta}=0)\ ,
\end{aligned}
\end{equation}
in the case in which $\Tilde{U}(0) = I$ and the blocks $\Tilde{U}$ are equal to each other. The orthogonality problem described above becomes more and more important when the number of qubits (i.e. the size of the Hilbert space) grows. Now, we know that $\ket{H\phi} = \sum_m P_m \ket{\phi}$, and we could try to engineer the empiric ansatz in a way such that the product of the vectors $\ket{T_i}$ and $\ket{P_m{\phi}}$ is not vanishing. The most straightforward way to do this is to impose that $\partial_m\ket{\phi}=iP_m\ket{\phi}$, but this is exactly the  VHA ansatz introduced above. \\
In fact thee VHA ansatz does not suffer from the problems analysed. First of all,  it is equal by construction to the identity when $\bar{\theta}=0$. Furthermore, it can be proven (see Appendix B) that when adopting this variational form, the VQS solution coincides with the exact propagator for vanishing small times, with parameters
\begin{equation}
    \theta_m = c_mt + \mathcal{O}(t^2) \ ,
\end{equation}
and thus respects exactly the condition III. This happens for \textit{any} state $\ket{\phi_0}$ that is evolved. In many cases of interest, the combination of this ansatz with the VQS algorithm allows to reduce remarkably the circuit depth with respect to the Suzuki-Trotter expansion technique. The fact that the VHA is a natural choice for the time evolution of states especially when the size of the Hilbert space grows must be taken into account when designing algorithms for near-term quantum computers and in the following we leverage on its properties to propose a new and optimized algorithm for the calculation of Green's functions.

\begin{figure}
	\centering
    \begin{subfigure}{0.5\textwidth}
        \includegraphics[scale=0.4]{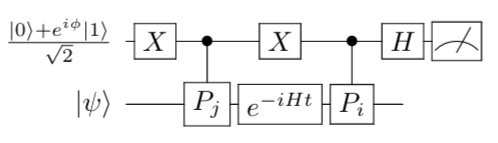}
        \caption{Circuit without CU}
        \label{no_CU}
    \end{subfigure}
    \begin{subfigure}{0.5\textwidth}
        \includegraphics[scale= 0.32]{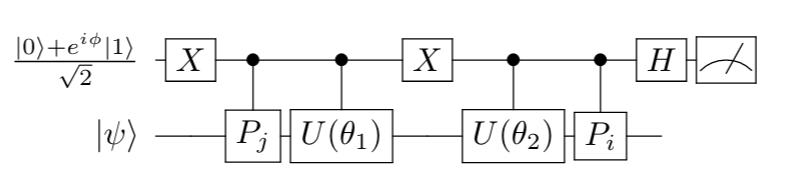}
        \caption{Circuit requiring CU}
        \label{yes_CU}
    \end{subfigure}
	\caption{Panel (a) shows the circuit for the calculation of correlation functions  previously suggested in literature. In Ref.~\cite{PhysRevX.6.031045,Kreula2016,Chiesa2019} the unitary operator evolving the system is calculated through Suzuki-Trotter expansion. Ref.~\cite{PhysRevResearch.2.033281}, instead, express the propagator $e^{-iHt}$ through a variational form $U(\bar{\theta})$ whose parameters are determined through the VQS McLachlan algorithm. It is important to stress that such approach works only if $U(\bar{\theta})$ approximates accurately the time evolution of the states $\ket{\psi}$ and $P_j\ket{\psi}$; otherwise it becomes necessary to adopt the quantum circuit shown in panel (b), which requires a controlled $U(\bar{\theta})$. }
	\label{standard}
\end{figure}

\section{Optimized algorithm}\label{IV}
\begin{figure}[h]
    \centering
        \begin{subfigure}{0.5\textwidth}
        \includegraphics[scale=0.35]{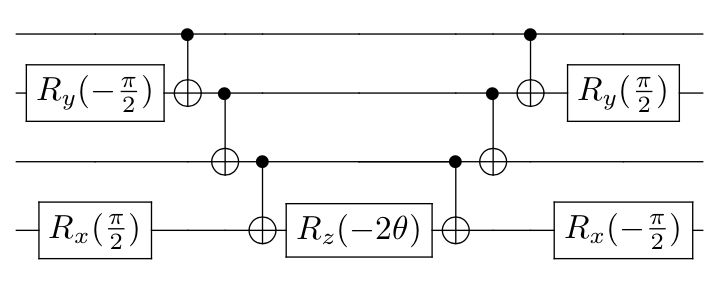}
        \caption{}
        \label{exp}
    \end{subfigure}
    \begin{subfigure}{0.5\textwidth}
        \includegraphics[scale= 0.35]{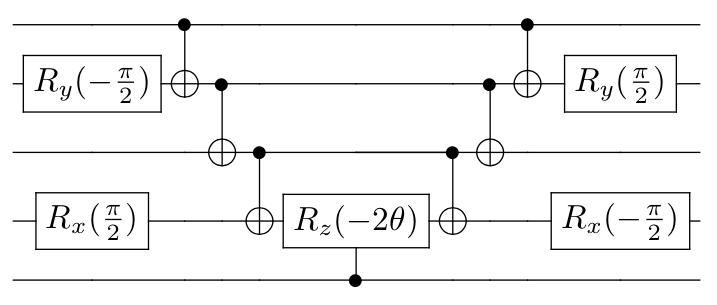}
        \caption{}
        \label{cexp}
    \end{subfigure}
    \caption{The circuit expressing the unitary $e^{-i\theta P_i}$ can be constructed using the rules of Ref.~\cite{https://doi.org/10.1002/qute.201900052}. Panel (a)  shows an example of the circuit corresponding to the exponential of the four-qubit Pauli gate ZXZY (namely $e^{-i\theta ZXZY})$. Panel (b) shows its controlled form. Notice that the controlled form is realized by simply controlling the central Rz gate alone. As a result, the controlled VHA simply requires the addition of a control operation for each appearing Pauli term, which is minimal considering that the exponential of a $n$-qubit Pauli gates requires at most $2 (n-1)$ CNOTS. }
    \label{fig:my_label}
\end{figure}

\begin{figure}[h]
    \centering
    \begin{subfigure}{0.5\textwidth}
        \includegraphics[scale=0.3]{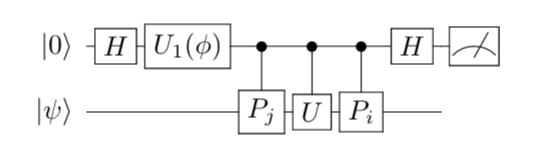}
        \caption{}
        \label{had}
    \end{subfigure}
    \begin{subfigure}{0.5\textwidth}
        \includegraphics[scale=0.3]{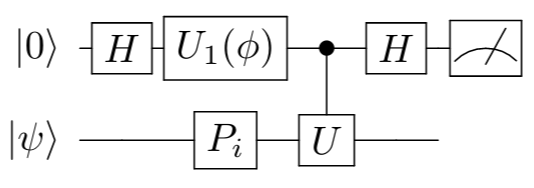}
        \caption{}
        \label{had_1}
    \end{subfigure}
    \caption{Panel (a) shows the circuit to calculate the braket of Eq. \ref{gf2}. When $\phi=0$, the Z-measurement of the first qubit gives the real part of the braket, while the imaginary part can be obtained by setting $\phi=-\frac{\pi}{2}$. The controlled operations on the Pauli gates $P_i$ and $P_j$ can be further removed if $P_i=P_j$, as shown in panel (b). }
    \label{hadamard_test}
\end{figure}

The standard technique to calculate correlation functions~\cite{PhysRevX.6.031045,Kreula2016,Chiesa2019} adopts the circuit shown in Fig~\ref{no_CU}, where the propagator $e^{-iHt}$ is calculated using the Suzuki-Trotter expansion. 
%\textcolor{red}{In my opinion the figure requires more explanations that will also help understanding the text below, cU, ...}.
Endo et al.~\cite{PhysRevResearch.2.033281} improved this algorithm by replacing the Trotter expansion, which requires very deep circuits, with a variational form $U$ whose parameters are determined through the VQS algorithm. In order for this to work, the operator $U$ has to accurately approximate the time evolution for the ground state $\ket{\psi}$  and the state $\ket{P_i\psi}$ at the same time, i.e. with the same parameters $\theta_i$ (the evolution of the ground state consists simply in adding the correct phase without modifying  the state). This requirement allows one to eliminate the controlled variational form $CU$ from the circuit (see Fig. \ref{yes_CU}), which would be otherwise necessary when $U$ evolves $\ket{\psi}$ and $\ket{P_i\psi}$ with a different choice of the parameters for the two states. For this reason, we shall refer to this algorithm as control-free (CF).
%The algorithm proposed by Endo et al. \cite{PhysRevResearch.2.033281} is based on the assumption that it is more convenient to eliminate the controlled unitary operation describing the evolution of the system at the price of requiring a ansatz which evolves at the same time the ground state $|\psi\rangle$ and the state $|P_i\psi\rangle$. 
Requiring $U$ to evolve two states at the same time seems to be a good price to pay to remove the $CU$ operations, since a controlled unitary operation acting on $n$ qubits is in general very complicated to realize and can require a large number of CNOTs for its implementation. However, it turns out that this strategy is not always advantageous:  first of all, as discussed in the previous section, it is difficult to find a shallow empiric variational form that can propagate a desired state with a sufficiently small error. This task is even harder as the size of the Hilbert space grows, particularly if one requires the ansatz to evolve two states simultaneously. This leads naturally to the choice of the VHA ansatz. Second, we argue that it is possible to control the VHA by adding a minimal number of controlled operations, as shown in Fig. \ref{cexp}. %Therefore avoiding the c-VHA does not lead to a significant advantage in term of controlled operations and circuit depth.
Therefore, the use of c-VHA does not necessarily imply a significant increase of controlled operations and, consequently, unaffordable circuit depths.
 %Conversely, 
Finally, evolving two states at the same time requires a deeper variational ansatz, with a larger cost in terms of controlled operations and circuit depth, as it will be proven numerically in the following sections. This is because the variational ansatz must approximate the ideal time evolution operator in at least two independent directions in the Hilbert space. \\
It is easy to conclude that an algorithm for the Green's function which requires a controlled evolution operator evolving one single state at a time can lead to an improvement in term of circuit depth and number of CNOTs.
This improved algorithm can be obtained under the assumption of a time independent Hamiltonian, which holds in many problems studied in the literature. In this case, the propagator is $e^{-iHt}$, and its action on the ground state introduces simply a phase rotation $e^{-iHt}|\psi\rangle = e^{-iE_0t}|\psi\rangle$, $E_0$ being the ground state energy which can be calculated through the variational quantum eigensolver (VQE) algorithm. Using this relation, the expression of the lesser Green's function becomes
\begin{equation}
    G_{lm}^<(t) = e^{iE_0t} \langle \psi| c_le^{-iHt} c_m^\dag |\psi \rangle\ .
\end{equation}
After mapping the fermionic Hamiltonian on a qubit register, the expression for the Green's function becomes
\begin{equation}\label{gf2}
    G_k^<(t) = e^{iE_0t}  \sum_{ij} \langle \psi| P_i U P_j^\dag |\psi \rangle  \lambda_i\lambda_j^* \,
\end{equation}
where $U$ is the ansatz of the VQS algorithm, and $\lambda_i$ are the coefficients of the expansion of the creation/destruction operators in term of Pauli gates. Since the operator $ P_i U P_j^\dag$ is not Hermitian, the braket must be computed through the Hadamard test; the circuit is the one shown in Fig.~\ref{standard}. It can be easily observed that in this case the ansatz only implies the evolution of the state $P_i|\psi\rangle$ at the price of a controlled evolution operator. Furthermore, when $P_i = P_j$, it is possible to remove the control operation on these gates, as shown in Fig.~\ref{had_1}.
Since the operator $U^{\dag}$ is removed from the expression, it is necessary to take correctly into account the global phase of the state $U\ket{P_j\psi}$. This can be done in a very simple way, as suggested in Ref.~\cite{Yuan2019theoryofvariational}, without the need to add a global phase to the ansatz used for the time evolution: indeed, the time dependent global phase $\theta_0(t)$ ensuring that the state $e^{i\theta_0(t)}U(\theta_1(t),...,\theta_m(t))\ket{\phi_0}$ coincides with the time evolution of the state $\ket{\phi_0}$ can be determined using the McLachlan principle. By manipulating Eq.~\ref{mclachlan1} we obtain the following differential equation:
\begin{equation}
    \dot{\theta_0} = \sum_{i=1}^M\Im{\braket{\partial_i \phi|\phi}}\ \dot{\theta_i} - \braket{\phi|H|\phi}\ ,
\end{equation}
where the $\dot{\theta_i}$ for $i=1,\dots, m$ are those calculated through the VQS method. In order to obtain the correct brakets, it is necessary to multiply the result of the circuit in Fig. \ref{hadamard_test} by $e^{i\theta_0(t)}$. 

The very same procedure can be followed to calculate the greater component of $G$, and thus the retarded correlation function. Since the proposed algorithm requires the evolution of one state at time, we shall refer to this algorithm as one-state (OS). In the next sections we will show the advantage of this algorithm for the 2- and 3-site Hubbard model with open-boundary conditions and 4-site with periodic-boundary conditions.

\section{2-site Hubbard model}\label{V}
\begin{figure}
    \centering
    \includegraphics[scale=0.5]{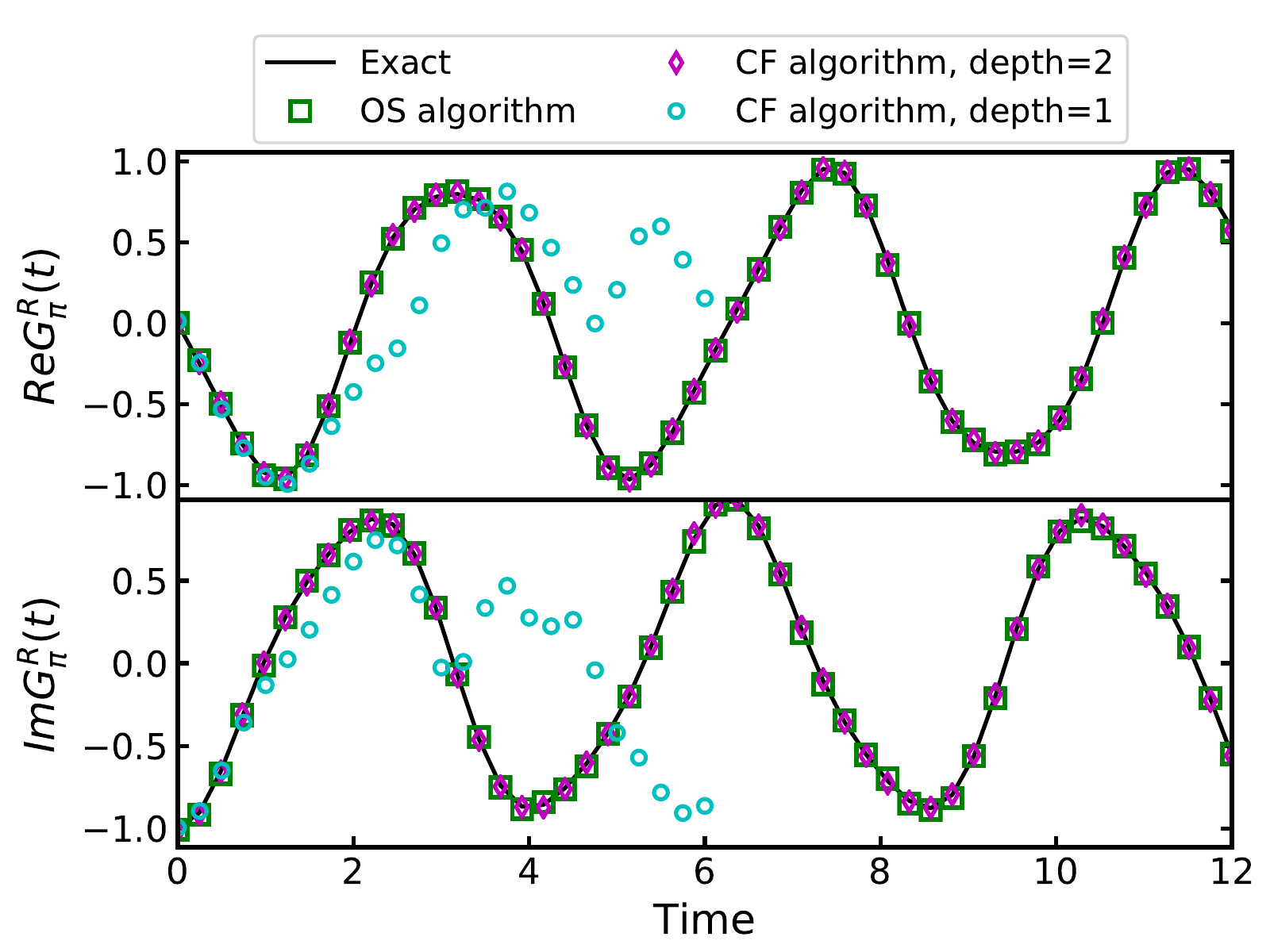}
    \caption{Real and imaginary part of the retarded Green's function $G^R$ calculated at $k=\pi\uparrow$. The black line corresponds to the exact Green's function, the green squares to the Green's function obtained through the OS algorithm, and the light blue circles and the violet diamonds to those calculated through the CF algorithm with depth equal to 1 and 2 respectively.}
    \label{gr_2}
\end{figure}

First we test our algorithm on the 2-site Hubbard model. It is possible to take advantage of the many symmetries of this model in order to reduce the depth of the ansatz used to evolve a generic state $P_i\ket{\psi}$. 
In particular, here we prove that the \textit{exact evolution} of the state $P_i\ket{\psi}$ is obtained by simply exponentiating one single Pauli component of the Hamiltonian (i.e. one among XZXI, YZYI, IXZX, IYZY). 
It is important to note that the present algorithm can exploit these peculiar symmetry properties, thanks to the fact that it only requires to implement the time-evolution of a single state $P_i\ket{\psi}$.  
Conversely, the CF algorithm, where the VHA ansatz has to evolve the states $P_i\ket{\psi}$ and $\ket{\psi}$ with the same parameters, requires not only the exponentiation of all Pauli terms in the Hamiltonian, but also the doubling of the circuit depth (depth=2), resulting in much deeper quantum circuits (see Fig.~\ref{gr_2}). \\ % and discussed later. \\
We start introducing the Hubbard dimer Hamiltonian
\begin{equation}\label{2-Hubbar}
\begin{aligned}
    H = -\tau \sum_{\sigma=\uparrow,\downarrow} (c_{1\sigma}^\dag c_{2\sigma} + c^\dag_{2\sigma} c_{1\sigma})
    + U\sum_{i=1,2} c_{i\uparrow}^\dag c_{i\uparrow}c_{i\downarrow}^\dag c_{i\downarrow}\\
    - \frac{U}{2} \sum_{i=1,2 \sigma= \uparrow,\downarrow} c_{i\sigma}^\dag c_{i\sigma}\ .
\end{aligned}
\end{equation}
The Jordan-Wigner transformation allow us to map the fermionic Hamiltonian into the following qubit Hamiltonian 
\begin{equation}
\begin{aligned}
\mathrm{
    H = -\frac{U}{2} IIII - \frac{\tau}{2} \Bigl( IXZX + IYZY + XZXI + YZYI \Bigr)}\\ 
    \mathrm{
    + \frac{U}{4} \Bigl( IIZZ + ZZII \Bigr) 
    } \ .
\end{aligned}
\end{equation}
Here we use the mapping $(1\uparrow,1\downarrow,2\uparrow,2\downarrow) \rightarrow  (q_1,q_2,q_3,q_4)$.
It is now possible to recognize two sets of operators, 
$$\mathrm{S_1 = \{ IXZX , IYZY , XZXI , YZYI \}}$$
and 
$$S_2 =\mathrm{ \{ ZZII, IIZZ \}}$$
such that, $\forall\ P_l,P_m \in S_i$:
\begin{enumerate}[label=(\roman*)]
    \item $[P_l, P_m] = 0$ (each pair of operators commute); \label{prop1}
    \item $[H, P_lP_m] = 0$ (the product of each pairs of operators commute with H) \label{prop2};
    \item $P_l^2=I$ \label{prop3}.
\end{enumerate}
These symmetries have very important consequences on the choices of the ansatz for the time evolution.
We will derive now a series of properties which will lead to drastic simplification of the time evolution circuit for the propagation of the Green's functions. \\
%simplify dramatically, the time evolution required for the Green's function.\\

\textbf{Proposition 1}.\textit{ If the ground state $|\psi\rangle$ of the Hamiltonian is non-degenerate, then $\forall\ P_l,P_m \in S_i \Longrightarrow$  $P_lP_m|\psi\rangle = |\psi\rangle.  $}\\
\indent
\textbf{Proof:} By definition, the ground state is the vector $|\psi\rangle$ such that $\langle\psi|H|\psi\rangle$ is minimum. However, $\langle\psi|H|\psi\rangle = \langle\psi|P^{\dag}_mP^{\dag}_l H P_lP_m|\psi\rangle$ from property (ii) above. Since we assume that the ground state is non-degenerate, then $P_lP_m|\psi\rangle = \alpha |\psi\rangle$. Using the commutation of $P_l$ and $P_m$ (i) we have also that $P_mP_l|\psi\rangle = \alpha |\psi\rangle$.
Now, $P_lP_mP_mP_l|\psi\rangle = \alpha^2 |\psi\rangle$. At the same time, since Pauli matrices square to unity, $P_lP_mP_mP_l|\psi\rangle =|\psi\rangle$; it follows that $\alpha=\pm 1$.
Finally, $\braket{\psi|P_l|\psi} + \braket{\psi|P_m|\psi} = \braket{\psi|P_mP_lP_m|\psi} + \braket{\psi|P_lP_mP_l|\psi} = \alpha(\braket{\psi|P_l|\psi} + \braket{\psi|P_m|\psi})$; therefore, $\alpha=1$.\\

\noindent
Following Proposition 1, it is possible to derive the following corollary.\\

\textbf{Proposition 2}.\textit{ If the ground state $|\psi\rangle$ of the Hamiltonian is non-degenerate, then $\forall\ P_l,P_m \in S_i \Longrightarrow$  $P_l|\psi\rangle = P_m|\psi\rangle.  $}\\
This result can be obtained by multiplying both the sides of the result of Proposition 1 by $P_l$ and using the property (iii) above.\\

Let us now proceed with the analysis of the time evolution of a generic state, which is obtained by applying the creation/destruction operators to the ground state $\ket{\psi}$. 
Because of the Jordan-Wigner transformation, this corresponds to applying a Pauli operator $\Tilde{P}$ (e.g. ZZXI) to the ground state; a tilde is used to remark that $\Tilde{P}\notin S_i$.\\
Let us consider $P_l,P_m \in S_i$. 
For Proposition 2, we have that $P_l\ket{\psi} = P_m\ket{\psi} = \ket{\chi}$. 
Therefore, 
$$P_l\Tilde{P}\ket{\psi} = \Tilde{P}\ket{\chi} + [P_l,\Tilde{P}]\ket{\psi} \,$$
and 
$$P_m\Tilde{P}\ket{\psi} = \Tilde{P}\ket{\chi} + [P_m,\Tilde{P}]\ket{\psi} \, .$$ 
In Appendix C we prove that only two cases are possible:
\begin{enumerate}
    \item $[P_l,\Tilde{P}]=0$, leading to  $P_l\Tilde{P}\ket{\psi} = \Tilde{P}\ket{\chi}$;
    \item $[P_l,\Tilde{P}] = 2P_l\Tilde{P}\ket{\psi} $, leading to $P_l\Tilde{P}\ket{\psi} = -\Tilde{P}\ket{\chi}$ \,.
\end{enumerate}
Using the simple rules described in Appendix C, we can then build the following table showing the sign of the term $P_l\Tilde{P}\ket{\psi}$ (i.e. $\pm\Tilde{P}\ket{\chi}$) for each couple of $P_l$ (rows) and $\Tilde{P}$ (columns).
\begin{table}[h]
    \centering
    \begin{tabular}{c|c|c|c|c|c}
         & XIII & ZXII & ZZXI & ZZZX & coeff.\\
        \hline\hline
        XZXI & + & + & - & + & -$\tau$/2\\
        YZYI & - & + & + & + & -$\tau$/2\\
        IXZX & + & + & + & - & -$\tau$/2\\
        IYZY & + & - & + & + & -$\tau$/2\\
        \hline
        ZZII & - & - & + & + & U/4\\
        IIZZ & + & + & - & -& U/4\\
    \end{tabular}
    \caption{Sign of the prefactor multiplying $\Tilde{P}\ket{\chi}$. The Pauli matrices in the left column are component of the qubit Hamiltonian, with coefficient reported in the last column, while the Pauli matrices in top row are the correspond to the fermion creation/annihilation operator in the qubit Hilbert space. }
    \label{sign}
\end{table}
It is now possible to relate these results with the time evolution of the operator $\Tilde{P}\ket{\psi}$. At first order in the expansion in t, 
\begin{equation}
    e^{-iHt}\Tilde{P}\ket{\psi} \simeq I -i\sum_m c_m t P_m \Tilde{P}\ket{\psi} \, .
\end{equation}
As the coefficients are the same for all the operators belonging to $S_i$, then the contributions of the elements of $S_2$ always cancel each other, while that of the elements of $S_1$ is always $2\Tilde{P}\ket{\chi}$, so
\begin{equation}
    e^{-iHt}\Tilde{P}\ket{\psi} \simeq I -i\ 2 c_1 t P_1 \Tilde{P}\ket{\psi} \, ,
\end{equation}
where $P_1$ is one of the Pauli matrices belonging to $S_1$.
For the second order expansion in t, 
\begin{equation}
    H^2\Tilde{P}\ket{\psi} = \sum_{lm} c_lc_m P_lP_m \Tilde{P}\ket{\psi}=2c_1 \sum_{l} c_l P_lP_1\Tilde{P}\ket{\psi} \ .
\end{equation}
Since the Pauli operator belonging to $S_2$ anticommute with all those belonging to $S_1$,
\begin{equation}\label{second}
    H^2\Tilde{P}\ket{\psi} = 2c_1 \sum_{l\in S_1} c_l P_1P_l\Tilde{P}\ket{\psi} - 2c_1 \sum_{l\in S_2} c_l P_1P_l\Tilde{P}\ket{\psi}
\end{equation}
The second term in the r.h.s. of Eq. \ref{second} vanishes as shown above, while the first term is equal to $4c_1^2 P_1^2$. It is straightforward to generalize the result at any order in perturbation theory, and conclude that the exact evolution of $\Tilde{P}\ket{\psi}$ is obtained by applying the operator
\begin{equation}\label{layer}
    U = e^{i \sigma\tau P_l t}
\end{equation}
where $P_l\in S_1$ and the sign $\sigma$ can be read in Tab.~\ref{sign} for each specific operator $\Tilde{P}$.

We make now use of this compact form of the propagator in the simulation of the two-site Hubbard model.
Fig.~\ref{gr_2}\ shows the results for the retarded Green's function calculated numerically with the Qiskit \textit{qasm} simulator~\cite{Qiskit}. The Green's function obtained with the OS algorithm is represented by green squares. 
Thanks to the properties shown above, the ansatz adopted is the one of Eq.~\ref{layer}, which contains only 5 CNOTs (in its controlled form). The Green's function obtained with the CF algorithm is represented with light blue circles (d=1) and violet diamonds (d=2). The former contains 20 CNOTs and fails to reproduce the exact Green's function dynamics, while the latter is in good agreement with it, but requires 40 CNOTs, as shown in Tab.~\ref{cnots}.
\begin{table*}[t]
    \centering
    \begin{tabular}{c|c|c|c|}
           \textbf{Two-site Hubbard model}\, &\ 1-qubit gates \,\, &\ 2-qubit gates \,\, &\ ansatz depth\,\, \\
           \hline
        OS algorithm & 4 & 5 & 9\\
        CF algorithm, d=1 \ & 22 & 20 & 42\\
        CF algorithm, d=2 \ & 44 & 40 & 84\\
        \textbf{3-site Hubbard model}\, &\ 1-qubit gates \,\, &\ 2-qubit gates \,\, &\ ansatz depth\,\, \\
           \hline
        OS algorithm, d=3 & 96 & 147 & 243\\
        CF algorithm, d=3 \ & 129 & 114 & 243\\
        CF algorithm, d=5 \ & 160 & 190 & 350\\
        \textbf{4-site Hubbard model}\, &\ 1-qubit gates \,\, &\ 2-qubit gates \,\, &\ ansatz depth\,\, \\
           \hline
        OS algorithm, d=3 & 192 & 372 & 564\\
        CF algorithm, d=4 \ & 336 & 416 & 752\\
        CF algorithm, d=5 \ & 420 & 520 & 940\\
    \end{tabular}
    \caption{Number of 1-qubit and 2-qubit gates in the ans\"atze adopted to calculate the Green's function for the Hubbard model. The number reported for the OS algorithm corresponds to the controlled form of the ansatz, which is the one adopted to calculate the Green's function.}
    \label{cnots}
\end{table*}
In summary, the novel OS algorithm manages to reach the same accuracy as the CF does; however, with a number of CNOTs and a total gate count of about one ninth.

\section{Scaling to larger problem sizes}\label{VI}
In order to provide further evidence about the advantages of the OS algorithm, we also performed calculations of the Green's function for the 3-site Hubbard chain with open-boundary conditions and the 4-site Hubbard chain with periodic-boundary conditions. 
In these cases, it is not possible to exploit the symmetries of the Hamiltonian to simplify the ansatz as much as for the 2-site Hubbard model. 
Nonetheless, we show that the suggested algorithm allows us to obtain an accurate Green's function with shallow circuits. % from Endo et al.

The 3-site Hubbard model is characterised by 2 degenerates ground states; here, we calculate the correlation function for one of these two ground states. The 4-site Hubbard model has instead a non-degenerate ground state. The retarded Green's function for both the models is computed at a $k$ value such that $c_k = c_{1\uparrow} - c_{2\uparrow}$. The result is shown in Fig.~\ref{gr_3sites}:
\begin{figure*}[t]
    \centering
    \begin{subfigure}{0.48\textwidth}
        \includegraphics[scale=0.5]{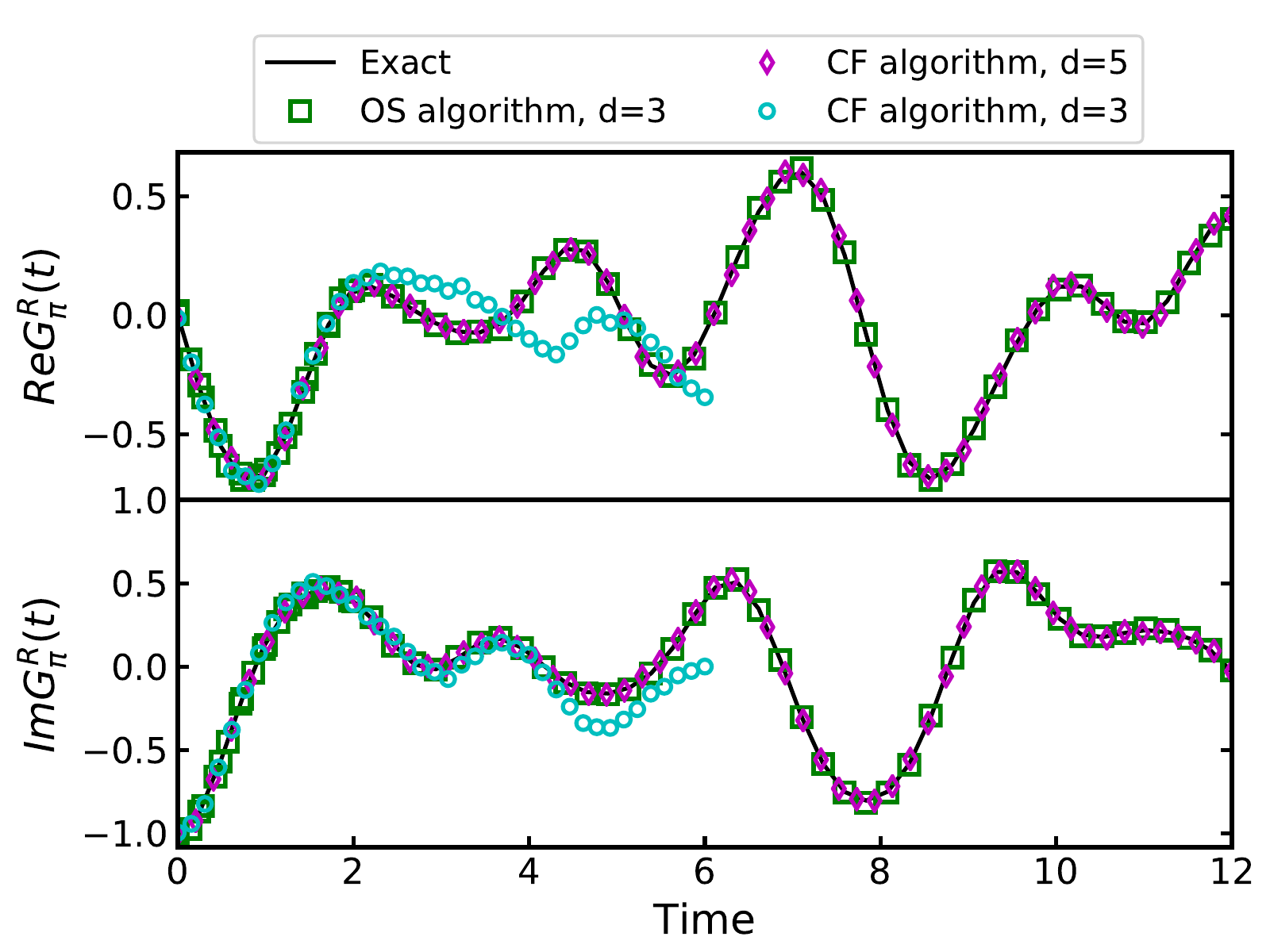}
        \caption{3-site Hubbard model.}
    \end{subfigure}
    \begin{subfigure}{0.48\textwidth}
        \includegraphics[scale=0.5]{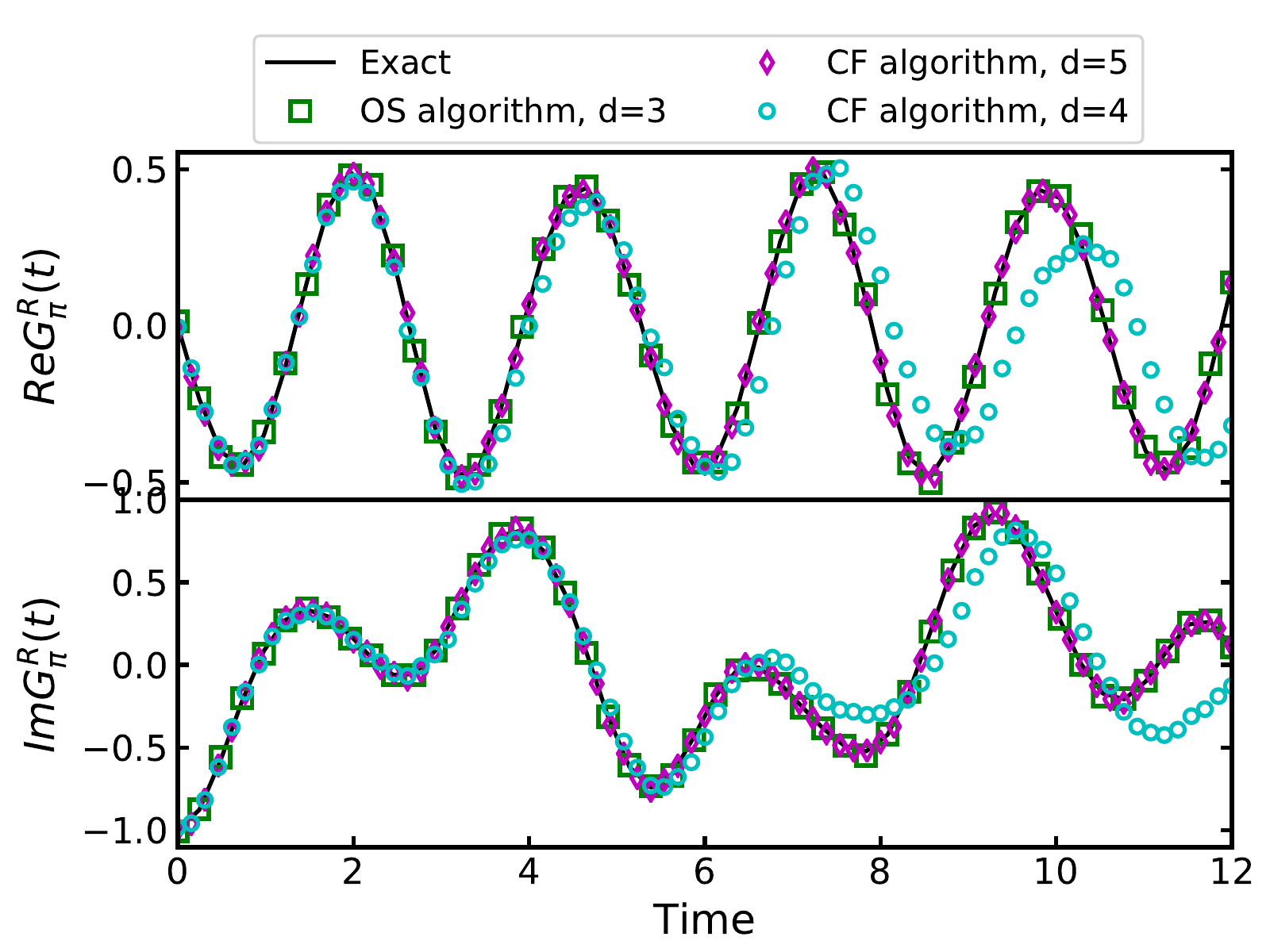}
        \caption{4-site Hubbard model.}
    \end{subfigure}
    \caption{Panel (a) shows the real and imaginary time retarded Green's function calculated at $c_k = c_{1\uparrow}-c_{2\uparrow}$ for the 3-site Hubbard model with the different algorithms analysed up to now. Since the ground state is degenerate, we consider the first diagonal component, which is obtained by sandwiching the operator $U^\dag c_k U c_k^\dag$ with one of the ground states $\ket{\psi_1}$ . Panel (b) shows the real and imaginary time retarded Green's function calculated at $c_k = c_{1\uparrow}-c_{2\uparrow}$ for the 4-site Hubbard model. }
    \label{gr_3sites}
\end{figure*}
we observe that the Green's function for the 3-site Hubbard model obtained with the new algorithm using a VHA of depth $d=3$ shows the same accuracy as the one computed with the CF algorithm with a depth $d=5$, which 
%The same algorithm 
fails to reproduce the exact Green's function when the depth is reduced to $d=3$. The same happens for the 4-site Hubbard model.
The advantage in term of number of CNOTs and circuit depth is shown in Tab.~\ref{cnots}.\\

To further investigate the scaling properties of the OS algorithm, let us summarize the main differences with respect to the CF protocol. On one hand, our strategy generally requires smaller VHA depths, thus reducing significantly the number of quantum operations necessary for a faithful description of the dynamics. On the other hand, one additional ancilla-controlled operation is necessary per Pauli string appearing in the Hamiltonian.
Under the decomposition scheme presented in Fig.~\ref{exp} -- employing the native gate set of current IBM Quantum processors --  the exponentiation of a single Pauli component of the Hamiltonian requires $\mathrm{2(\mathit{w}-1)}$ CNOTs, where $w$ is the corresponding Pauli weight, i.e.~the number of non-identity elements in the Pauli string.
The total number of controlled operations required by the OS algorithm for a VHA of depth $d_1$ is thus:
\begin{equation}\label{n1}
n^{CNOT}_1 = d_1\sum_{i=1}^{n_p} (2(w_i-1)+1)=d_1n_p\ (2(\bar{w}-1)+1)\ ,\end{equation}
where $n_p$ is the number of Pauli components of the Hamiltonian, $w_i$ is the weight of the i-th component and $\bar{w}$ is the average Pauli weight, defined as $\bar{w}=1/n_p\sum_{i=1}^{n_p} w_i\ .$
The CF algorithm requires instead, for a VHA of depth $d_2$
\begin{equation}\label{n2}
n^{CNOT}_2 = d_2\sum_{i=1}^{n_p} 2(w_i-1) = d_2 n_p\ 2(\bar{w}-1)\ .\end{equation}
Combining Eqs. \ref{n1} and \ref{n2} it is easy to see that the suggested algorithm will show an advantage over the CF one when
\begin{equation}\label{fin}
    \frac{d_2}{d_1} > 1 + \frac{1}{2(\bar{w}-1)}\ .
\end{equation}
Clearly, the OS algorithm is favoured for larger average Pauli weights. In this respect, the n-site Hubbard chain with open-boundary conditions and first nearest neighbours coupling represents the least favourable case: indeed, the average Pauli weight $\bar{w}=(14n-12)/(5n-4)$ has an asymptotic value of 2.8, thus requiring $d_2/d_1>1.27$ for the OS algorithm to be competitive. However, when replacing the open-boundary conditions with periodic-boundary conditions, or when assuming that the Hubbard sites are disposed on a 2d lattice instead of a chain, the appearance of Pauli components of maximum Pauli weight makes the condition of Eq.~\ref{fin} much more favourable (see Appendix D). Finally, in a typical quantum chemistry problem, the interacting term $h_{pqrs}a_{p}^\dag a_q^{\dag}a_ra_s$ is mapped into 
\begin{equation}
\begin{aligned}
    H_{pqrs} = \frac{h_{pqrs}}{8} \prod_{j=s+1}^{r-1}Z_j\prod_{k=q+1}^{p-1}Z_k \qquad\quad\\
    (XXXX-XXYY+XYXY+YXXY+\\
    YXYX-YYXX+XYYX+YYYY)\ ,
\end{aligned}
\end{equation}
which can easily lead to large Pauli weights. 
%Therefore we expect our algorithm to work better for complex problems. 

It is worth reminding that, while all the examples above were treated by making use of the Jordan-Wigner fermion-to-qubit mapping, other strategies are known, such as the Bravyi-Kitaev mapping~\cite{BRAVYI2002210}, which in general leads to lower average Pauli weight in the asymptotic limit. These alternative mappings also have an impact on the VHA implementation details and required depths, thus requiring a case-by-case assessment~\cite{tranter2018}.

Finally, it is also important to remark that the controlled VHA requires in general a higher connectivity between the ancilla and the system qubits. As a result, native hardware coupling maps and circuit compilation protocols must also be taken into account when selecting the the most favorable strategy for specific applications.

\section{Hardware results}\label{VII}
The algorithm presented in this paper allows one to reduce remarkably the size of the circuit used to compute  Green's functions for several model Hamiltonians. 
In the case of the 2-site Hubbard model, the improvement is such that it allows the execution of the algorithm on a current quantum processor. More specifically, we demonstrate an implementation of our OS algorithm on IBM Quantum superconducting devices, accessible via the cloud~\footnote{See \url{https://quantum-computing.ibm.com/.}}.

\begin{figure*}[t]
    \centering
    \includegraphics[scale = 0.5]{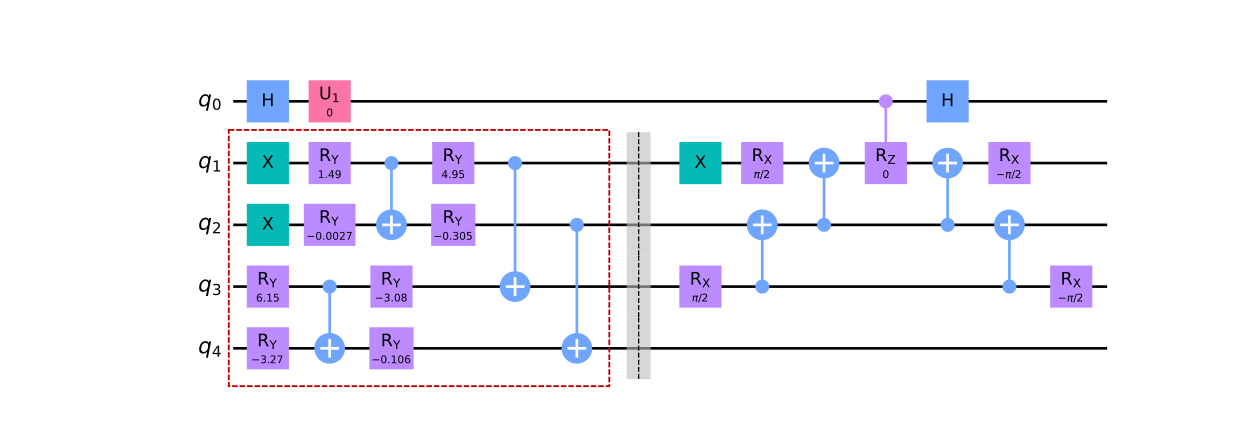}
    \caption{Circuit required to calculate the braket in Eq. \ref{gf2} when $\mathrm{P_i=P_j=XIII}$ at $t=0$. The part of the circuit which is enclosed in the red rectangle corresponds to the variational form adopted for the VQE. }
    \label{circ_tot}
\end{figure*}

We first obtain an approximate ground state using the VQE algorithm with the SPSA optimizer on \textit{ibmq\_manila}. 
The adopted variational form is represented in Fig.~\ref{circ_tot} (within the red box), and consists of a concatenation of single $R_y$ qubit rotations and CNOTs.
For this experiment, we used a Hubbard Hamiltonian  with $t=1$ and $U=3$. 
The theoretical ground state energy corresponds to $E_0=-4$, while the experimental energy obtained with the VQE is -3.63. 
This result can be further improved on \textit{ibmq\_montreal}, due to the lower noise level of the processor. By also applying readout error mitigation~\cite{PRXQuantum.2.040326}, we obtain a ground state energy of~$-3.91$. 

\begin{figure*}[t]
    \centering
    \begin{subfigure}{0.48\textwidth}
        \includegraphics[scale=0.5]{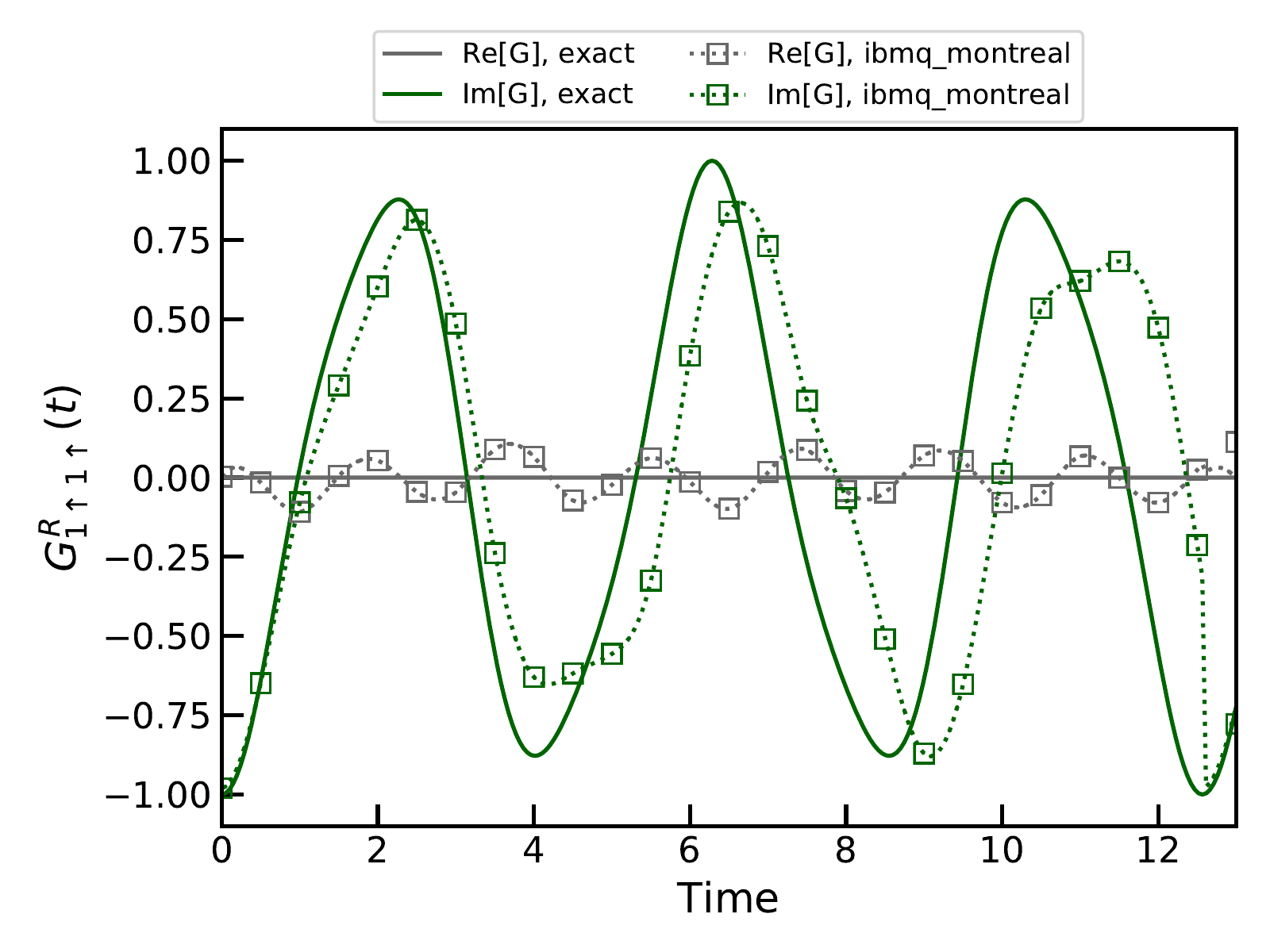}
        \caption{Time domain, noisy $\mathrm{\Tilde{E}_0}$.}
        \label{res_3}
    \end{subfigure}
    \begin{subfigure}{0.48\textwidth}
        \includegraphics[scale=0.5]{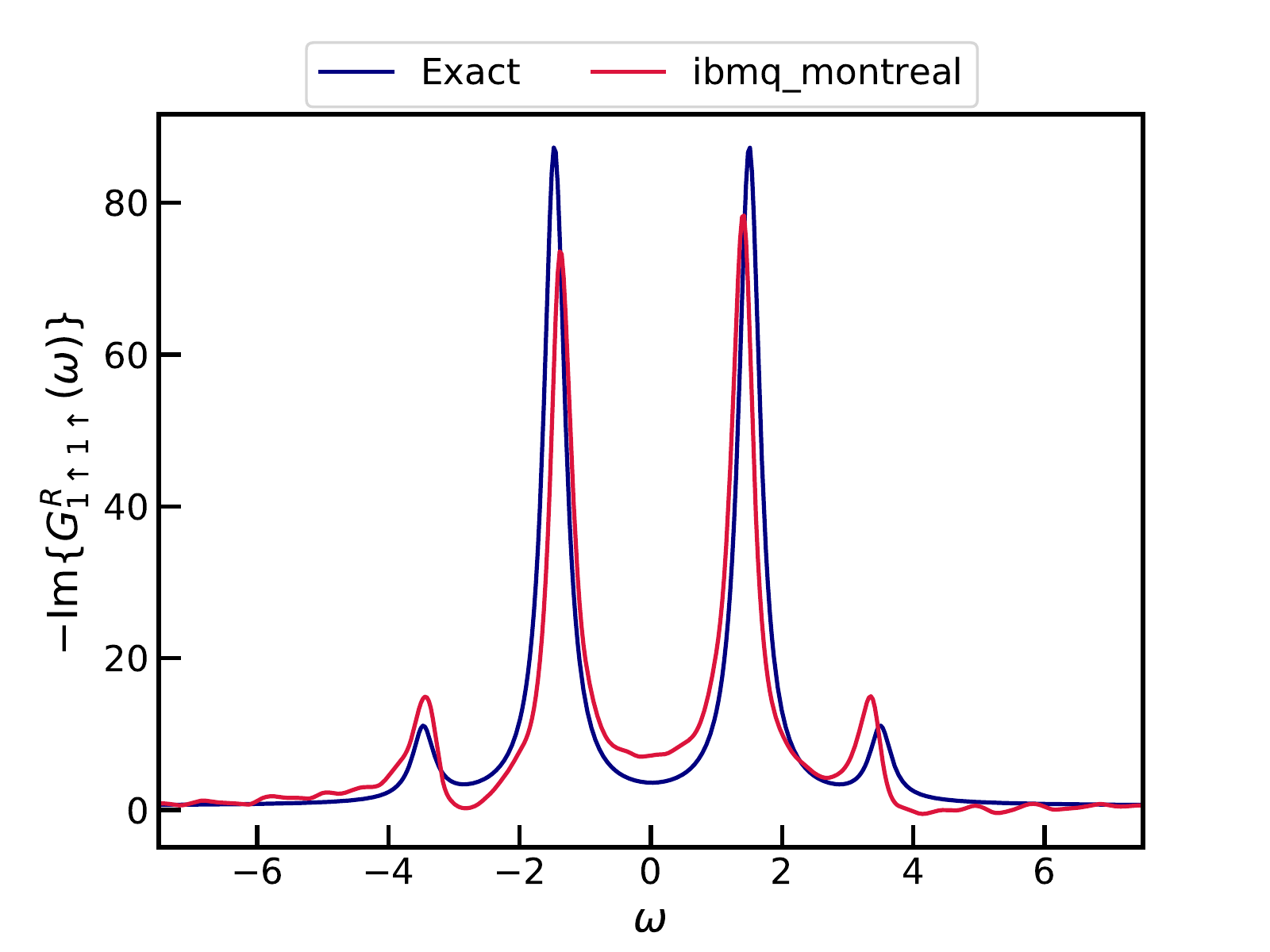}
        \caption{Frequency domain, noisy $\mathrm{\Tilde{E}_0}$.}
        \label{res_2}
    \end{subfigure}    
    \begin{subfigure}{0.48\textwidth}
        \includegraphics[scale=0.5]{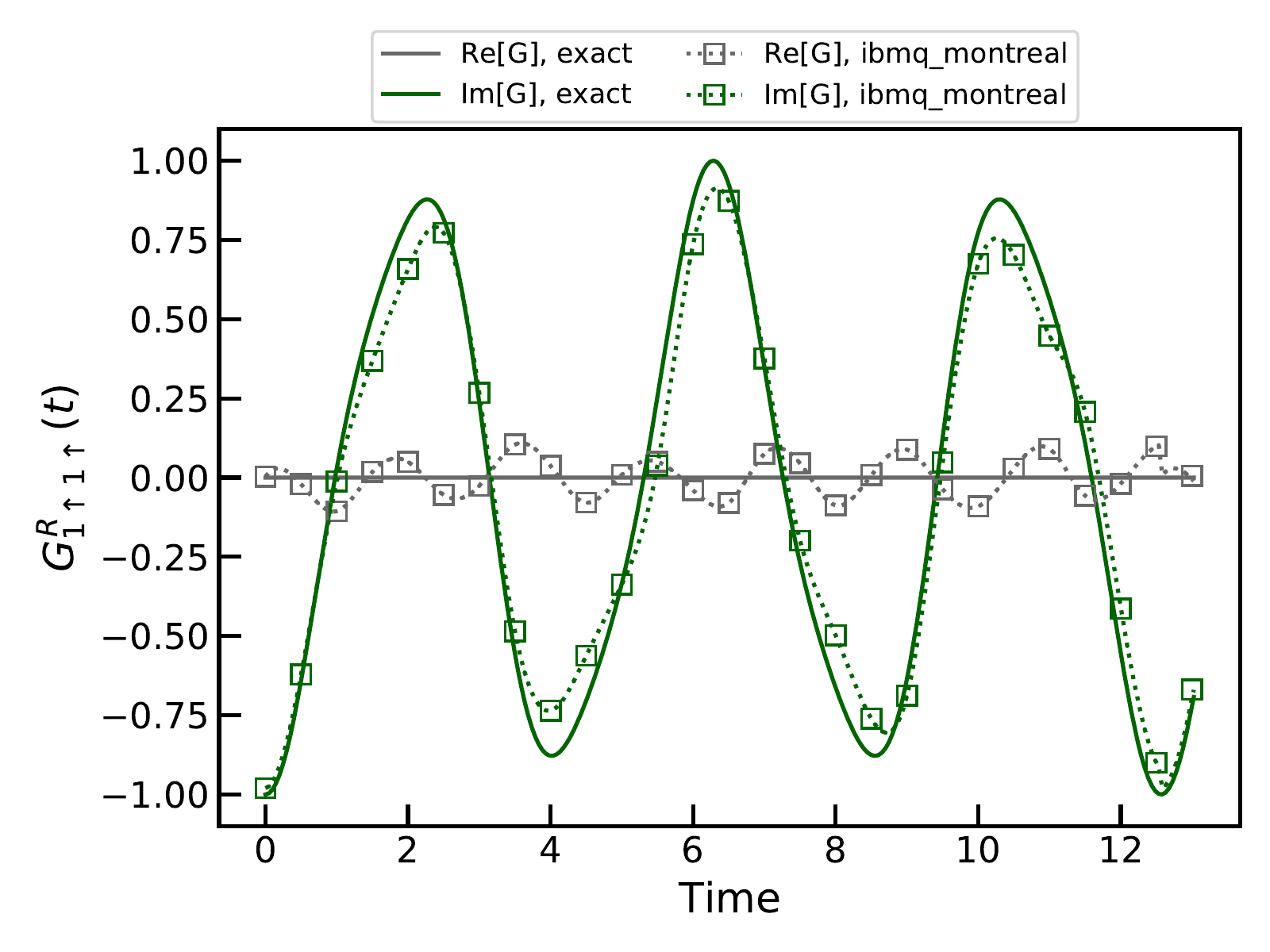}
        \caption{Time domain, exact $\mathrm{E_0}$.}
        \label{res_1}
    \end{subfigure}
    \begin{subfigure}{0.48\textwidth}
        \includegraphics[scale=0.5]{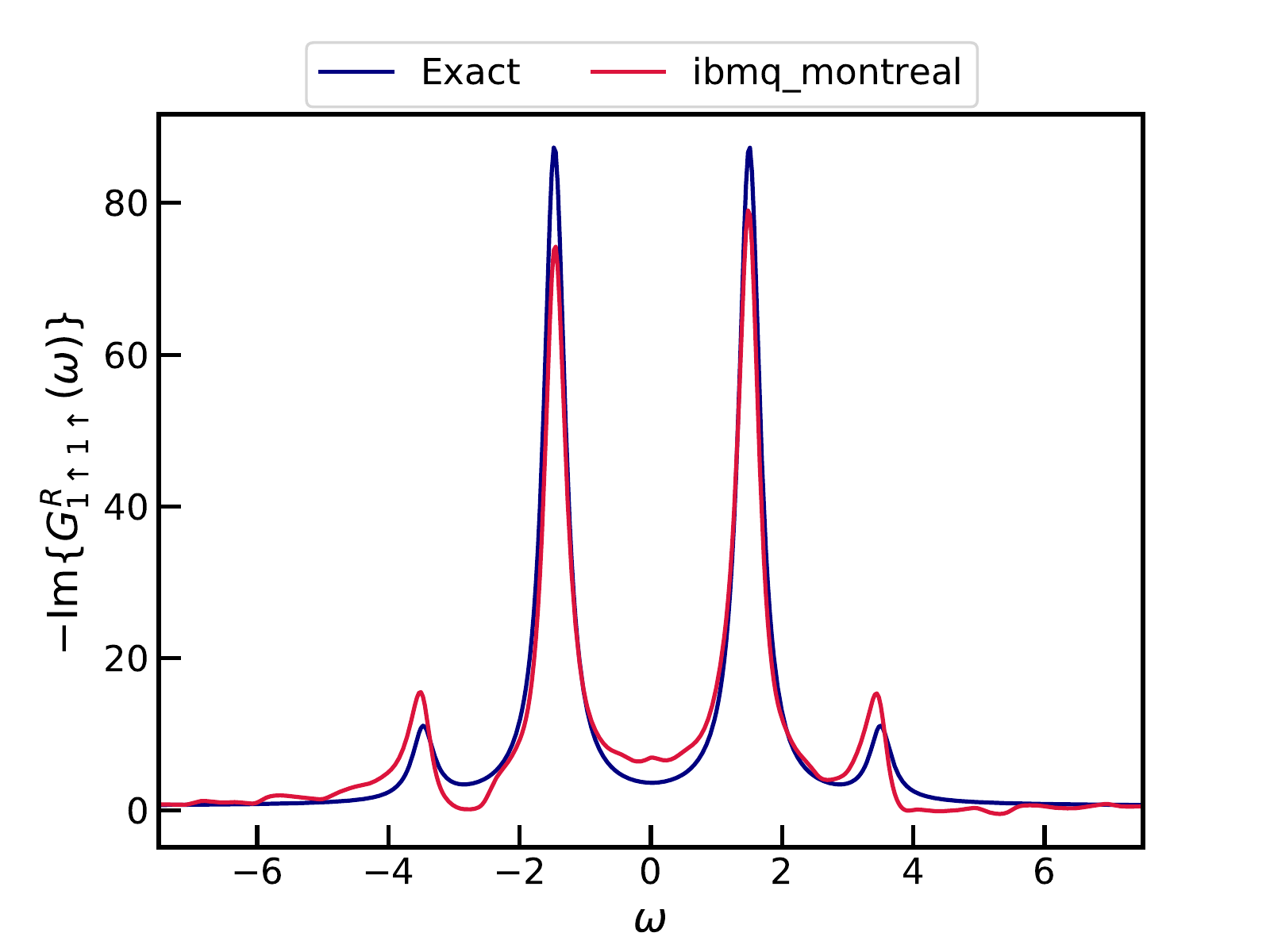}
        \caption{Frequency domain, exact $\mathrm{E_0}$.}
        \label{res_4}
    \end{subfigure}
    \caption{Panel (a) shows the retarded Green's function in time domain. The grey and green solid lines corresponds to the exact real and imaginary parts respectively. The grey and green squares correspond to the results from the hardware. The dotted lines are obtained by interpolating the hardware result with splines. Panel (b) shows the imaginary part of the retarded Green's function in frequency domain. The blue and the red solid lines correspond respectively to the exact result and the quantum calculation. Panels (c) and (d) show the same quantities as Panels (a) and (b), with the only difference that now the Green's function is calculated using the exact energy $\mathrm{E}_0=-4$ in the phase factor $e^{iE_0t}$ of Eq. \ref{gf2}. }
    \label{hardware_result}
\end{figure*}

After the optimization, we calculate the diagonal component of the retarded Green's function, $G_{1\uparrow 1\uparrow}^R$. 
The full circuit used for the calculation of the expectation values in Eq.~\ref{gf2} is given in Fig~\ref{circ_tot} for the case of $P_i=P_j=XIII$ at $t=0$; further details on the calculation setup are given in Appendix E. 
Due to the particle-hole symmetry, its real part is vanishing and only the imaginary part survives. 
The calculation is performed on \textit{ibmq\_montreal}, and the results are shown in Fig.~\ref{res_3}. This figure, however is not meaningful for a visual comparison with the experiments, due to the frequency shift introduced by the error on $E_0$. It is instead much more instructive to consider the Green's function obtained from Eq.~\ref{gf2}, 
where all matrix elements are measured on the quantum processor while the value of the energy in the phase of $e^{iE_0t}$ is taken to be equal to its theoretical value, $\mathrm{E_0=-4}$ (instead of taking the approximated variational energy computed on \textit{ibmq\_montreal}) in Fig. \ref{res_1}.
%by calculating the sum over i and j with the quantum computer and using the theoretical $\mathrm{E_0=-4}$ value instead of the noisy $\mathrm{\Tilde{E}_0=}-3.91$ in the phase $e^{iE_0t}$. 
The reasons are threefold.  
%First of all, the adoption of the  energy $\mathrm{\Tilde{E}_0}$ would introduce a small frequency shift that does not allow to appreciate visually the accuracy of the algorithm for the evaluation of the expectation values in Eq.~\ref{gf2}. 
First, the adoption of the  energy $\mathrm{E_0}$ eliminates the small frequency shift and allows to appreciate visually the accuracy of the algorithm for the evaluation of the brakets in Eq.~\ref{gf2}.
Second, the ground state energy is just a parameter in Eq. \ref{gf2} that could be evaluated with higher accuracy using other techniques (such as imaginary time evolution). 
Finally and most importantly, a small error in the ground state energy only leads to a small energy shift of the poles of the Green's function in frequency domain, as it can be seen by comparing Fig. \ref{res_2} with Fig. \ref{res_4}, where the noisy energy $\mathrm{\Tilde{E}_0}$ and the exact energy $\mathrm{E_0}$ are used respectively (the proof is given in Appendix E). This agrees with the general observation that, thorugh e.g.~Fourier analysis, useful and accurate information in the frequency domain can be obtained even from noisy quantum data~\cite{Chiesa2019,neill_accurately_2021}. 
Overall, the agreement shown in Fig. \ref{res_1} between the imaginary part of the exact Green's function and the one calculated on the hardware is very good, also considering that no further error mitigation scheme is adopted at this stage. 
The real part of the Green's function obtained on hardware shows small oscillations around the exact value 0. These oscillations are mainly caused by the errors of the hardware in reproducing the true ground state wavefunction. 
In the frequency domain, the agreement between the exact result and the quantum computation is overall very good, both for the shape and the position of the peaks, also when the noisy energy $\mathrm{\Tilde{E}_0}$ is adopted. 

\section{Conclusions}
In this work, we proposed an optimized real time propagation algorithm for the calculation of the Green's function on a near-term quantum computer, which exploits the time independence of the Hamiltonian and the features of the VHA ansatz and requires the time evolution of the only state $P_i\ket{\psi}$.
We proved that this algorithm is more efficient in term of 2-qubit gates than the previously known ones, both for the 2-site 3-site and 4-site Hubbard models, and we estimates that the reported advantage is likely to persist when considering complex realistic quantum chemistry problems.
In particular, this algorithm is able exploit the symmetries of the 2-site Hubbard model, where the exact propagator of the state $P_i\ket{\psi}$ is obtained by exponentiating only one of the Pauli components of the Hamiltonian. This leads to a very shallow circuit for the calculation of the Green's function which, can then be executed on a current quantum computer, with results in remarkable agreement with the exact solutions. \\
We studied the scaling of the algorithm with the size of the problems, and proved that the algorithm proposed is favoured in problems where the Pauli weight of the components of the Hamiltonian is large. This is the case, for example, of quantum chemistry problems with interactions beyond first nearest neighbours. \\
We believe that the present work will promote  further research towards the development of hardware efficient quantum algorithms for calculation of electronic correlation functions, opening up new avenues in the study of complex many-body systems and their excitations on near-term quantum computers.
%\textcolor{red}{Needs some extentions: 1. beyond the 2 side Hubbard in the last paragraph, 2. projections to other systems in chemistry and material science, i.e. add comments about chemistry Hamiltonian.}

\section{Acknowledgments}

This research was supported by an E$^3$ (EPFL Excellence
in Engineering) program fellowship, the grants
200021-179312, 200021-179138 and NCCR MARVEL,
funded by the Swiss National Science Foundation.
IBM, the IBM logo, and ibm.com are trademarks
of International Business Machines Corp., registered in
many jurisdictions worldwide. Other product and service
names might be trademarks of IBM or other companies.
The current list of IBM trademarks is available at
https://www.ibm.com/legal/copytrade.

\onecolumngrid
\appendix

\section{Geometry of McLachlan principle}
In this section we prove Eq. \ref{geometric} of the main text. 
First, we use the identity
\begin{equation}
    \braket{\phi|\phi} = 1 \qquad \forall t
\end{equation}
to show that $\braket{\partial_i\phi|\phi}$ is a pure imaginary number. In fact
\begin{equation}
    \partial_i \braket{\phi|\phi} = 0 = \braket{\partial_i \phi|\phi} + \braket{\phi|\partial_i \phi} \Longrightarrow \braket{\phi|\partial_i \phi} = -\braket{\phi|\partial_i \phi}^*
\end{equation}
Therefore 
\begin{equation}
    \braket{\partial_i \phi|\phi} \braket{\partial_j \phi|\phi} = 
    - \braket{\partial_i \phi|\phi} \braket{\phi| \partial_j \phi} =
    - \Re{\braket{\partial_i \phi|\phi} \braket{\phi| \partial_j \phi}}
\end{equation}
Substituting in Eq. \ref{M} of the main text, 
$$
    M_{ij} = \Re{\Bigl[ \braket{\partial_i\phi|\partial_j\phi} - \braket{\partial_i \phi|\phi} \braket{\phi| \partial_j \phi}} \Bigr] =  \Re{\braket{T_i|T_j}}\ .
$$
Similarly, since $\braket{\phi|H|\phi}$ is real, then
\begin{equation}
    i \braket{\partial_i\phi|\phi}\braket{\phi|H|\phi} = 
    -\Im{\braket{\partial_i\phi|\phi}\braket{\phi|H|\phi}}\ .
\end{equation}
Substituting in Eq. \ref{Vi} of the main text, 
\begin{equation}
     V_{i} = \Im \Bigl[ {\braket{\partial_i\phi|H|\phi} - \braket{\partial_i\phi|\phi}\braket{\phi|H|\phi} } \Bigr]  = \braket{T_i|H|\phi}
\end{equation}

\section{}
In this section we show that, when using the VHA anzatz (Eq. \ref{VHA}), McLachlan-VQS solution coincide with the exact evolution for vanishingly small times. At t=0, the partial derivative of the VHA with respect to the parameter $\theta_m$ is
\begin{equation}
    |\partial_m \phi \rangle \Bigl|_{t=0}  = -i P_m \ket{\phi_0} \ ;
\end{equation}
this means that
\begin{equation}
    H |\phi_0 \rangle   = \sum_{m} c_m P_m \ket{\phi_0}  = \sum_m i c_m  |\partial_m \phi \rangle \ .
\end{equation}
Therefore 
\begin{equation}
    \langle T_l | H |\phi_0 \rangle  = \sum_m i c_m  \langle T_l |\partial_m \phi \rangle  =  \sum_m i c_m  \langle T_l |T_m \rangle \ .
\end{equation}
If the $c_m$ are reals, then
\begin{equation}\label{eq1}
    \Im[V_l] = \sum_m c_m\Re[\braket{T_l |T_m}]\ .
\end{equation}
Substituting Eq. \ref{eq1} in Eq. \ref{geometric} of the main text, we obtain 
\begin{equation}
    \Re[\braket{T_l |T_m}] \dot{\theta}_m = c_m\Re[\braket{T_l |T_m}]
\end{equation}
which admit as solution 
\begin{equation}
     \dot{\theta}_m = c_m\ ,
\end{equation}
therefore
\begin{equation}
     \theta_m(t) = c_mt + o(t^2) ,
\end{equation}

\section{Algebric properties of Pauli matrices}
Let us consider two n-dimensional Pauli operators, $P,P'$. We can write (without restriction) 
$$P = P_1P_2...P_mP_{m+1}...P_n$$
$$P' = P'_1P'_2...P'_mP'_{m+1}...P'_n$$
where the first corresponding m operators of $P$ and $P'$ are Pauli operators which are different from each other and from identity (e.g $P_1 = X$,\ $P'_1=Y$) and the corresponding operators from $m+1$ to $n$ are either equal to each other or one of them is equal to the identity (e.g $P_n = I$,\ $P'_n=Y$ or $P_n = Y$,\ $P'_n=Y$).
We have that
\begin{equation}
PP' = (i\varepsilon_{11'1''}P_1'')...(i\varepsilon_{mm'm''}P_m'')(P_{m+1}P_{m+1}'...P_{n}P_{n}')
\end{equation}
\begin{equation}
\begin{aligned}
[P,P'] = (i\varepsilon_{11'1''}P_1'')...(i\varepsilon_{mm'm''}P_m'')(P_{m+1}P_{m+1}'...P_{n}P_{n}') \\
- (-i\varepsilon_{11'1''}P_1'')...(-i\varepsilon_{mm'm''}P_m'')(P_{m+1}P_{m+1}'...P_{n}P_{n}')
\end{aligned}
\end{equation}
$$\Longrightarrow [P,P'] = PP'(1-(-1)^m)$$
So, if $m$ is even, then the commutator is vanishes.
If, $m$ is odd, then $[P,P'] = 2PP'$.

\section{Pauli weight of Hubbard models}
The n-site Hubbard chain with first-nearest neighbours hopping and time reversal symmetry has the following Hamiltonian:
\begin{equation}
\begin{aligned}
H = -\tau\sum_{i=1}^{n-1}\sum_{\sigma=\uparrow,\downarrow}\ (c_{i,\sigma}^\dag c_{i+1,\sigma}+c_{i+1,\sigma} c_{i,\sigma}^\dag ) \\
+ U\sum_{i=1}^{n} c_{i\uparrow}^\dag c_{i\uparrow}c_{i\downarrow}^\dag c_{i\downarrow}
    - \frac{U}{2} \sum_{i=1}^n \sum_{\sigma= \uparrow,\downarrow} c_{i\sigma}^\dag c_{i\sigma}\ ,
\end{aligned}
\end{equation}
where $n$ is the number of sites.
The Wigner-Jordan transformation maps this Hamiltonian into the following set of 2n-qubit Pauli gates, denoted by $H_n$
%\begin{equation}
%    H_n = \Bigl\{\mathrm{XZXI...I , YZYI...I, IXZXI...I, IYZYI...I, ..., I...IYZY} \Bigl\} \bigcup 
%    \Bigl\{ \mathrm{ZZII...II, IIZZII...II, ..., II...IIZZ }  \Bigl\} 
%\end{equation}
\begin{equation}
    H_n = H_n^{(1)} \bigcup H_n^{(2)}\ .
\end{equation}
The elements of $H_n^{(1)}$ have the form 
\begin{equation}
    e_{x} = \mathrm{III...XZX...II}  
\end{equation}
and 
\begin{equation}
    e_{y} = \mathrm{III...YZY...II} \ ,
\end{equation}
where the triplet XZX and YZY can assume any position in the qubit register. The number of elements of this set are thus 4(n-1), and their weight is $w=3$. 
The elements of $H_n^{(2)}$, instead, have the form  
\begin{equation}
    e_{z} = \mathrm{II..ZZ...II} \ ,
\end{equation}
whit an even number of identities I before and/or after the doublet ZZ. The number of elements of this set are thus n, and their weight is $w=2$.
The average Pauli weight is
\begin{equation}
    \bar{w} = \frac{4(n-1)\times 3+n\times 2}{4(n-1)+n} = \frac{14n-12}{5n-4}
\end{equation}
whose asymptotic value is $\bar{w}$=2.8.
When replacing open-boundary conditions with periodic-boundary condition, the hopping between the first and the n-th site cause the appearance of terms such as $e_x=\mathrm{ZXZZZ...ZX}$, where the Z gates act on all the central qubits, so the weight is maximum $w=2n$. The same condition holds when the Hubbard sites are not displaced on a row (Hubbard chain) but on a 2d or even on a 3d lattice, increasing remarkably the average Pauli weight of the problem.

\section{Details about the calculation of the Green's function}
\begin{figure}[h]
    \centering
    \includegraphics[scale=0.5]{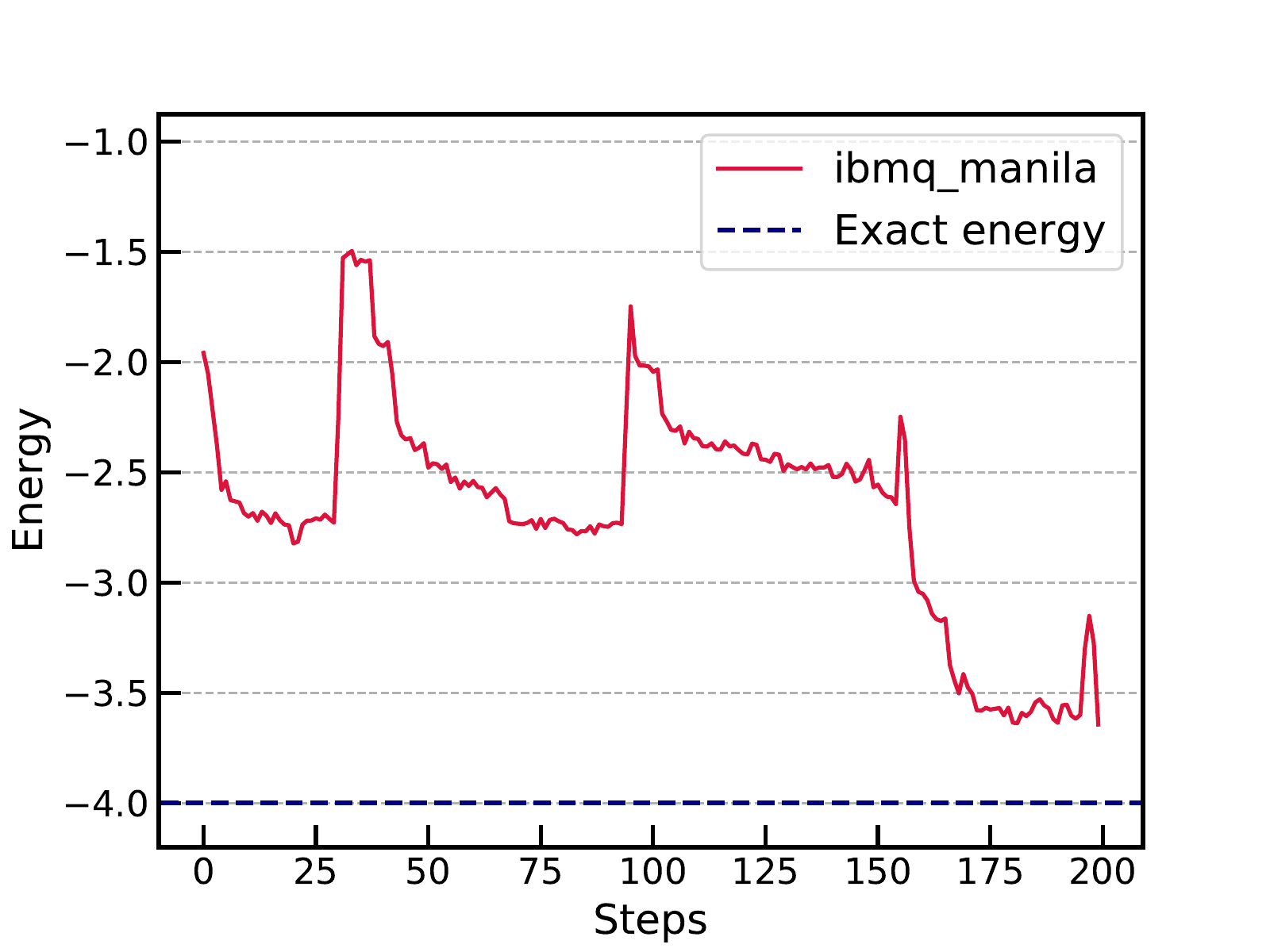}
    \caption{VQE algorithm for the ground state on \textit{ibmq\_manila}. The blue dashed line corresponds to the exact value, while the red line to the quantum calculation.}
    \label{vqe_opt}
\end{figure}
\begin{figure}
    \centering
    \includegraphics[scale=0.4]{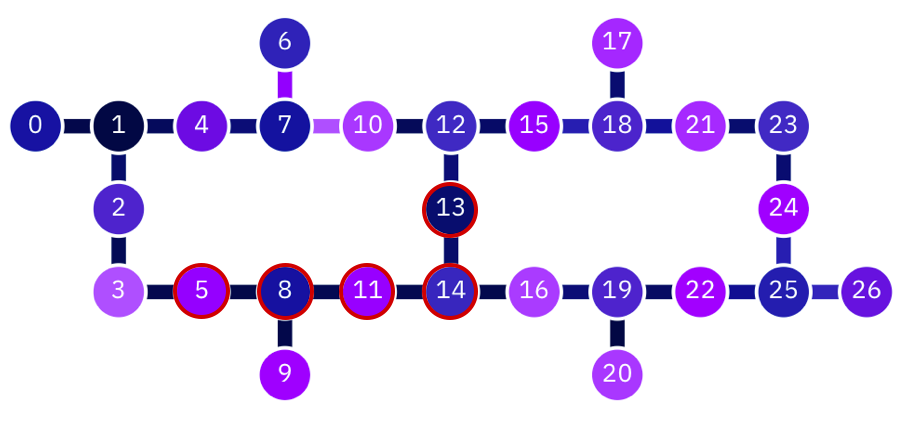}
    \caption{Layout of $ibmq\_montreal$. The qubits used in the calculation of the Green's function are those circled in red. }
    \label{montreal}
\end{figure}

The optimization of the ground state energy is reported in Fig. \ref{vqe_opt}.\\
The Green's function is calculated on \textit{ibmq\_montreal}. The layout of this quantum computer, together with the qubits used in the calculation are reported in Fig. \ref{montreal}.\\
As it is well known from the many-body perturbation theory \cite{Fetter}, the poles of the Green's function correspond to the energy of the charged excitations, $E_i(N\pm1)-E_0(N)$, where N is the number of particles in the ground state (in this case N=2) and $E_i(N\pm1)$ is the energy of the i-th excited states with $N\pm1$ particles. In the 2-site Hubbard model with parameters U=3 and t=1 there are four poles, corresponding to the energies $\Delta E_1 = \pm 1.5$ and $\Delta E_2=\pm3.5$ (see Ref. \cite{Stefanucci} for the analytic solution of the Hubbard model), therefore the Green's function has a periodicity $T=4\pi$. This is true only if the value of $E_0$ used in Eq. \ref{gf2} is the exact one, as the noisy value $\Tilde{E}_0$ would imply a longer period. In order to save quantum-computational time, we adopt the following trick (NB this trick, which relies on our knowledge of the exact solution of the problem, is definitely not necessary for the calculation of the Green's function, and it has been used here only to reduce the number of calculations ran on the quantum computer). First of all, we write the lesser term 
\begin{equation}
    G_k^<(t) = -ie^{iE_0t} \langle \psi| c_ke^{-iHt} c_k^\dag |\psi \rangle =
    -i e^{i(\Tilde{E}_0-E_0)t} e^{iE_0t} \langle \psi| c_ke^{-iHt} c_k^\dag |\psi \rangle\ ,
\end{equation}
with $g^<(t)= -ie^{iE_0t} \langle \psi| c_ke^{-iHt} c_k^\dag |\psi \rangle\ $ periodic on $T=4\pi$. We therefore calculate $g(t)$ over this time range and then its Fourier transform $g(\omega)$. Thanks to the convolution theorem, 
\begin{equation}
    G_k^<(\omega) = g^<(\omega + \Tilde{E}_0 - E_0 )\ ,
\end{equation}
therefore we just need to shift the function $g(\omega)$ of $E_0 - \Tilde{E}_0$ in order to obtain the correct Fourier transform. We can repeat the very same procedure for the greater term and thus obtain the retarded component.
\twocolumngrid

\bibliography{main}

\begin{thebibliography}{32}%
\makeatletter
\providecommand \@ifxundefined [1]{%
 \@ifx{#1\undefined}
}%
\providecommand \@ifnum [1]{%
 \ifnum #1\expandafter \@firstoftwo
 \else \expandafter \@secondoftwo
 \fi
}%
\providecommand \@ifx [1]{%
 \ifx #1\expandafter \@firstoftwo
 \else \expandafter \@secondoftwo
 \fi
}%
\providecommand \natexlab [1]{#1}%
\providecommand \enquote  [1]{``#1''}%
\providecommand \bibnamefont  [1]{#1}%
\providecommand \bibfnamefont [1]{#1}%
\providecommand \citenamefont [1]{#1}%
\providecommand \href@noop [0]{\@secondoftwo}%
\providecommand \href [0]{\begingroup \@sanitize@url \@href}%
\providecommand \@href[1]{\@@startlink{#1}\@@href}%
\providecommand \@@href[1]{\endgroup#1\@@endlink}%
\providecommand \@sanitize@url [0]{\catcode `\\12\catcode `\$12\catcode
  `\&12\catcode `\#12\catcode `\^12\catcode `\_12\catcode `\%12\relax}%
\providecommand \@@startlink[1]{}%
\providecommand \@@endlink[0]{}%
\providecommand \url  [0]{\begingroup\@sanitize@url \@url }%
\providecommand \@url [1]{\endgroup\@href {#1}{\urlprefix }}%
\providecommand \urlprefix  [0]{URL }%
\providecommand \Eprint [0]{\href }%
\providecommand \doibase [0]{http://dx.doi.org/}%
\providecommand \selectlanguage [0]{\@gobble}%
\providecommand \bibinfo  [0]{\@secondoftwo}%
\providecommand \bibfield  [0]{\@secondoftwo}%
\providecommand \translation [1]{[#1]}%
\providecommand \BibitemOpen [0]{}%
\providecommand \bibitemStop [0]{}%
\providecommand \bibitemNoStop [0]{.\EOS\space}%
\providecommand \EOS [0]{\spacefactor3000\relax}%
\providecommand \BibitemShut  [1]{\csname bibitem#1\endcsname}%
\let\auto@bib@innerbib\@empty
%</preamble>
\bibitem [{\citenamefont {Golze}\ \emph {et~al.}(2019)\citenamefont {Golze},
  \citenamefont {Dvorak},\ and\ \citenamefont
  {Rinke}}]{10.3389/fchem.2019.00377}%
  \BibitemOpen
  \bibfield  {author} {\bibinfo {author} {\bibfnamefont {D.}~\bibnamefont
  {Golze}}, \bibinfo {author} {\bibfnamefont {M.}~\bibnamefont {Dvorak}}, \
  and\ \bibinfo {author} {\bibfnamefont {P.}~\bibnamefont {Rinke}},\ }\href
  {\doibase 10.3389/fchem.2019.00377} {\bibfield  {journal} {\bibinfo
  {journal} {Frontiers in Chemistry}\ }\textbf {\bibinfo {volume} {7}},\
  \bibinfo {pages} {377} (\bibinfo {year} {2019})}\BibitemShut {NoStop}%
\bibitem [{\citenamefont {Aryasetiawan}\ and\ \citenamefont
  {Gunnarsson}(1998)}]{Aryasetiawan_1998}%
  \BibitemOpen
  \bibfield  {author} {\bibinfo {author} {\bibfnamefont {F.}~\bibnamefont
  {Aryasetiawan}}\ and\ \bibinfo {author} {\bibfnamefont {O.}~\bibnamefont
  {Gunnarsson}},\ }\href {\doibase 10.1088/0034-4885/61/3/002} {\bibfield
  {journal} {\bibinfo  {journal} {Reports on Progress in Physics}\ }\textbf
  {\bibinfo {volume} {61}},\ \bibinfo {pages} {237} (\bibinfo {year}
  {1998})}\BibitemShut {NoStop}%
\bibitem [{\citenamefont {Ren}\ \emph {et~al.}(2015)\citenamefont {Ren},
  \citenamefont {Marom}, \citenamefont {Caruso}, \citenamefont {Scheffler},\
  and\ \citenamefont {Rinke}}]{PhysRevB.92.081104}%
  \BibitemOpen
  \bibfield  {author} {\bibinfo {author} {\bibfnamefont {X.}~\bibnamefont
  {Ren}}, \bibinfo {author} {\bibfnamefont {N.}~\bibnamefont {Marom}}, \bibinfo
  {author} {\bibfnamefont {F.}~\bibnamefont {Caruso}}, \bibinfo {author}
  {\bibfnamefont {M.}~\bibnamefont {Scheffler}}, \ and\ \bibinfo {author}
  {\bibfnamefont {P.}~\bibnamefont {Rinke}},\ }\href {\doibase
  10.1103/PhysRevB.92.081104} {\bibfield  {journal} {\bibinfo  {journal} {Phys.
  Rev. B}\ }\textbf {\bibinfo {volume} {92}},\ \bibinfo {pages} {081104}
  (\bibinfo {year} {2015})}\BibitemShut {NoStop}%
\bibitem [{\citenamefont {Georges}\ \emph {et~al.}(1996)\citenamefont
  {Georges}, \citenamefont {Kotliar}, \citenamefont {Krauth},\ and\
  \citenamefont {Rozenberg}}]{RevModPhys.68.13}%
  \BibitemOpen
  \bibfield  {author} {\bibinfo {author} {\bibfnamefont {A.}~\bibnamefont
  {Georges}}, \bibinfo {author} {\bibfnamefont {G.}~\bibnamefont {Kotliar}},
  \bibinfo {author} {\bibfnamefont {W.}~\bibnamefont {Krauth}}, \ and\ \bibinfo
  {author} {\bibfnamefont {M.~J.}\ \bibnamefont {Rozenberg}},\ }\href {\doibase
  10.1103/RevModPhys.68.13} {\bibfield  {journal} {\bibinfo  {journal} {Rev.
  Mod. Phys.}\ }\textbf {\bibinfo {volume} {68}},\ \bibinfo {pages} {13}
  (\bibinfo {year} {1996})}\BibitemShut {NoStop}%
\bibitem [{\citenamefont {Stefanucci}\ and\ \citenamefont {van
  Leeuwen}(2013)}]{Stefanucci}%
  \BibitemOpen
  \bibfield  {author} {\bibinfo {author} {\bibfnamefont {G.}~\bibnamefont
  {Stefanucci}}\ and\ \bibinfo {author} {\bibfnamefont {R.}~\bibnamefont {van
  Leeuwen}},\ }\href@noop {} {\emph {\bibinfo {title} {Nonequilibrium Many-Body
  theory of Quantum Systems}}}\ (\bibinfo  {publisher} {Cambridge University
  Press},\ \bibinfo {address} {Boston},\ \bibinfo {year} {2013})\BibitemShut
  {NoStop}%
\bibitem [{\citenamefont {Baker}(2021)}]{PhysRevA.103.032404}%
  \BibitemOpen
  \bibfield  {author} {\bibinfo {author} {\bibfnamefont {T.~E.}\ \bibnamefont
  {Baker}},\ }\href {\doibase 10.1103/PhysRevA.103.032404} {\bibfield
  {journal} {\bibinfo  {journal} {Phys. Rev. A}\ }\textbf {\bibinfo {volume}
  {103}},\ \bibinfo {pages} {032404} (\bibinfo {year} {2021})}\BibitemShut
  {NoStop}%
\bibitem [{\citenamefont {Wecker}\ \emph
  {et~al.}(2015{\natexlab{a}})\citenamefont {Wecker}, \citenamefont {Hastings},
  \citenamefont {Wiebe}, \citenamefont {Clark}, \citenamefont {Nayak},\ and\
  \citenamefont {Troyer}}]{PhysRevA.92.062318}%
  \BibitemOpen
  \bibfield  {author} {\bibinfo {author} {\bibfnamefont {D.}~\bibnamefont
  {Wecker}}, \bibinfo {author} {\bibfnamefont {M.~B.}\ \bibnamefont
  {Hastings}}, \bibinfo {author} {\bibfnamefont {N.}~\bibnamefont {Wiebe}},
  \bibinfo {author} {\bibfnamefont {B.~K.}\ \bibnamefont {Clark}}, \bibinfo
  {author} {\bibfnamefont {C.}~\bibnamefont {Nayak}}, \ and\ \bibinfo {author}
  {\bibfnamefont {M.}~\bibnamefont {Troyer}},\ }\href {\doibase
  10.1103/PhysRevA.92.062318} {\bibfield  {journal} {\bibinfo  {journal} {Phys.
  Rev. A}\ }\textbf {\bibinfo {volume} {92}},\ \bibinfo {pages} {062318}
  (\bibinfo {year} {2015}{\natexlab{a}})}\BibitemShut {NoStop}%
\bibitem [{\citenamefont {Roggero}\ and\ \citenamefont
  {Carlson}(2019)}]{PhysRevC.100.034610}%
  \BibitemOpen
  \bibfield  {author} {\bibinfo {author} {\bibfnamefont {A.}~\bibnamefont
  {Roggero}}\ and\ \bibinfo {author} {\bibfnamefont {J.}~\bibnamefont
  {Carlson}},\ }\href {\doibase 10.1103/PhysRevC.100.034610} {\bibfield
  {journal} {\bibinfo  {journal} {Phys. Rev. C}\ }\textbf {\bibinfo {volume}
  {100}},\ \bibinfo {pages} {034610} (\bibinfo {year} {2019})}\BibitemShut
  {NoStop}%
\bibitem [{\citenamefont {Kosugi}\ and\ \citenamefont
  {Matsushita}(2020)}]{PhysRevA.101.012330}%
  \BibitemOpen
  \bibfield  {author} {\bibinfo {author} {\bibfnamefont {T.}~\bibnamefont
  {Kosugi}}\ and\ \bibinfo {author} {\bibfnamefont {Y.-i.}\ \bibnamefont
  {Matsushita}},\ }\href {\doibase 10.1103/PhysRevA.101.012330} {\bibfield
  {journal} {\bibinfo  {journal} {Phys. Rev. A}\ }\textbf {\bibinfo {volume}
  {101}},\ \bibinfo {pages} {012330} (\bibinfo {year} {2020})}\BibitemShut
  {NoStop}%
\bibitem [{\citenamefont {Bauer}\ \emph {et~al.}(2016)\citenamefont {Bauer},
  \citenamefont {Wecker}, \citenamefont {Millis}, \citenamefont {Hastings},\
  and\ \citenamefont {Troyer}}]{PhysRevX.6.031045}%
  \BibitemOpen
  \bibfield  {author} {\bibinfo {author} {\bibfnamefont {B.}~\bibnamefont
  {Bauer}}, \bibinfo {author} {\bibfnamefont {D.}~\bibnamefont {Wecker}},
  \bibinfo {author} {\bibfnamefont {A.~J.}\ \bibnamefont {Millis}}, \bibinfo
  {author} {\bibfnamefont {M.~B.}\ \bibnamefont {Hastings}}, \ and\ \bibinfo
  {author} {\bibfnamefont {M.}~\bibnamefont {Troyer}},\ }\href {\doibase
  10.1103/PhysRevX.6.031045} {\bibfield  {journal} {\bibinfo  {journal} {Phys.
  Rev. X}\ }\textbf {\bibinfo {volume} {6}},\ \bibinfo {pages} {031045}
  (\bibinfo {year} {2016})}\BibitemShut {NoStop}%
\bibitem [{\citenamefont {Kreula}\ \emph {et~al.}(2016)\citenamefont {Kreula},
  \citenamefont {Clark},\ and\ \citenamefont {Jaksch}}]{Kreula2016}%
  \BibitemOpen
  \bibfield  {author} {\bibinfo {author} {\bibfnamefont {J.~M.}\ \bibnamefont
  {Kreula}}, \bibinfo {author} {\bibfnamefont {S.~R.}\ \bibnamefont {Clark}}, \
  and\ \bibinfo {author} {\bibfnamefont {D.}~\bibnamefont {Jaksch}},\ }\href
  {\doibase 10.1038/srep32940} {\bibfield  {journal} {\bibinfo  {journal}
  {Scientific Reports}\ }\textbf {\bibinfo {volume} {6}},\ \bibinfo {pages}
  {32940} (\bibinfo {year} {2016})}\BibitemShut {NoStop}%
\bibitem [{\citenamefont {Chiesa}\ \emph {et~al.}(2019)\citenamefont {Chiesa},
  \citenamefont {Tacchino}, \citenamefont {Grossi}, \citenamefont {Santini},
  \citenamefont {Tavernelli}, \citenamefont {Gerace},\ and\ \citenamefont
  {Carretta}}]{Chiesa2019}%
  \BibitemOpen
  \bibfield  {author} {\bibinfo {author} {\bibfnamefont {A.}~\bibnamefont
  {Chiesa}}, \bibinfo {author} {\bibfnamefont {F.}~\bibnamefont {Tacchino}},
  \bibinfo {author} {\bibfnamefont {M.}~\bibnamefont {Grossi}}, \bibinfo
  {author} {\bibfnamefont {P.}~\bibnamefont {Santini}}, \bibinfo {author}
  {\bibfnamefont {I.}~\bibnamefont {Tavernelli}}, \bibinfo {author}
  {\bibfnamefont {D.}~\bibnamefont {Gerace}}, \ and\ \bibinfo {author}
  {\bibfnamefont {S.}~\bibnamefont {Carretta}},\ }\href {\doibase
  10.1038/s41567-019-0437-4} {\bibfield  {journal} {\bibinfo  {journal} {Nature
  Physics}\ }\textbf {\bibinfo {volume} {15}},\ \bibinfo {pages} {455}
  (\bibinfo {year} {2019})}\BibitemShut {NoStop}%
\bibitem [{\citenamefont {Endo}\ \emph {et~al.}(2020)\citenamefont {Endo},
  \citenamefont {Kurata},\ and\ \citenamefont
  {Nakagawa}}]{PhysRevResearch.2.033281}%
  \BibitemOpen
  \bibfield  {author} {\bibinfo {author} {\bibfnamefont {S.}~\bibnamefont
  {Endo}}, \bibinfo {author} {\bibfnamefont {I.}~\bibnamefont {Kurata}}, \ and\
  \bibinfo {author} {\bibfnamefont {Y.~O.}\ \bibnamefont {Nakagawa}},\ }\href
  {\doibase 10.1103/PhysRevResearch.2.033281} {\bibfield  {journal} {\bibinfo
  {journal} {Phys. Rev. Research}\ }\textbf {\bibinfo {volume} {2}},\ \bibinfo
  {pages} {033281} (\bibinfo {year} {2020})}\BibitemShut {NoStop}%
\bibitem [{\citenamefont {Nakanishi}\ \emph {et~al.}(2019)\citenamefont
  {Nakanishi}, \citenamefont {Mitarai},\ and\ \citenamefont
  {Fujii}}]{PhysRevResearch.1.033062}%
  \BibitemOpen
  \bibfield  {author} {\bibinfo {author} {\bibfnamefont {K.~M.}\ \bibnamefont
  {Nakanishi}}, \bibinfo {author} {\bibfnamefont {K.}~\bibnamefont {Mitarai}},
  \ and\ \bibinfo {author} {\bibfnamefont {K.}~\bibnamefont {Fujii}},\ }\href
  {\doibase 10.1103/PhysRevResearch.1.033062} {\bibfield  {journal} {\bibinfo
  {journal} {Phys. Rev. Research}\ }\textbf {\bibinfo {volume} {1}},\ \bibinfo
  {pages} {033062} (\bibinfo {year} {2019})}\BibitemShut {NoStop}%
\bibitem [{\citenamefont {Rizzo}\ \emph {et~al.}(2022)\citenamefont {Rizzo},
  \citenamefont {Libbi}, \citenamefont {Tacchino}, \citenamefont {Ollitrault},
  \citenamefont {Marzari},\ and\ \citenamefont
  {Tavernelli}}]{rizzo2022oneparticle}%
  \BibitemOpen
  \bibfield  {author} {\bibinfo {author} {\bibfnamefont {J.}~\bibnamefont
  {Rizzo}}, \bibinfo {author} {\bibfnamefont {F.}~\bibnamefont {Libbi}},
  \bibinfo {author} {\bibfnamefont {F.}~\bibnamefont {Tacchino}}, \bibinfo
  {author} {\bibfnamefont {P.~J.}\ \bibnamefont {Ollitrault}}, \bibinfo
  {author} {\bibfnamefont {N.}~\bibnamefont {Marzari}}, \ and\ \bibinfo
  {author} {\bibfnamefont {I.}~\bibnamefont {Tavernelli}},\ }\href@noop {}
  {\enquote {\bibinfo {title} {One-particle green's functions from the quantum
  equation of motion algorithm},}\ } (\bibinfo {year} {2022}),\ \Eprint
  {http://arxiv.org/abs/2201.01826} {arXiv:2201.01826 [quant-ph]} \BibitemShut
  {NoStop}%
\bibitem [{\citenamefont {Ollitrault}\ \emph {et~al.}(2020)\citenamefont
  {Ollitrault}, \citenamefont {Kandala}, \citenamefont {Chen}, \citenamefont
  {Barkoutsos}, \citenamefont {Mezzacapo}, \citenamefont {Pistoia},
  \citenamefont {Sheldon}, \citenamefont {Woerner}, \citenamefont {Gambetta},\
  and\ \citenamefont {Tavernelli}}]{PhysRevResearch.2.043140}%
  \BibitemOpen
  \bibfield  {author} {\bibinfo {author} {\bibfnamefont {P.~J.}\ \bibnamefont
  {Ollitrault}}, \bibinfo {author} {\bibfnamefont {A.}~\bibnamefont {Kandala}},
  \bibinfo {author} {\bibfnamefont {C.-F.}\ \bibnamefont {Chen}}, \bibinfo
  {author} {\bibfnamefont {P.~K.}\ \bibnamefont {Barkoutsos}}, \bibinfo
  {author} {\bibfnamefont {A.}~\bibnamefont {Mezzacapo}}, \bibinfo {author}
  {\bibfnamefont {M.}~\bibnamefont {Pistoia}}, \bibinfo {author} {\bibfnamefont
  {S.}~\bibnamefont {Sheldon}}, \bibinfo {author} {\bibfnamefont
  {S.}~\bibnamefont {Woerner}}, \bibinfo {author} {\bibfnamefont {J.~M.}\
  \bibnamefont {Gambetta}}, \ and\ \bibinfo {author} {\bibfnamefont
  {I.}~\bibnamefont {Tavernelli}},\ }\href {\doibase
  10.1103/PhysRevResearch.2.043140} {\bibfield  {journal} {\bibinfo  {journal}
  {Phys. Rev. Research}\ }\textbf {\bibinfo {volume} {2}},\ \bibinfo {pages}
  {043140} (\bibinfo {year} {2020})}\BibitemShut {NoStop}%
\bibitem [{\citenamefont {Yuan}\ \emph {et~al.}(2019)\citenamefont {Yuan},
  \citenamefont {Endo}, \citenamefont {Zhao}, \citenamefont {Li},\ and\
  \citenamefont {Benjamin}}]{Yuan2019theoryofvariational}%
  \BibitemOpen
  \bibfield  {author} {\bibinfo {author} {\bibfnamefont {X.}~\bibnamefont
  {Yuan}}, \bibinfo {author} {\bibfnamefont {S.}~\bibnamefont {Endo}}, \bibinfo
  {author} {\bibfnamefont {Q.}~\bibnamefont {Zhao}}, \bibinfo {author}
  {\bibfnamefont {Y.}~\bibnamefont {Li}}, \ and\ \bibinfo {author}
  {\bibfnamefont {S.~C.}\ \bibnamefont {Benjamin}},\ }\href {\doibase
  10.22331/q-2019-10-07-191} {\bibfield  {journal} {\bibinfo  {journal}
  {{Quantum}}\ }\textbf {\bibinfo {volume} {3}},\ \bibinfo {pages} {191}
  (\bibinfo {year} {2019})}\BibitemShut {NoStop}%
\bibitem [{\citenamefont {Wecker}\ \emph
  {et~al.}(2015{\natexlab{b}})\citenamefont {Wecker}, \citenamefont
  {Hastings},\ and\ \citenamefont {Troyer}}]{PhysRevA.92.042303}%
  \BibitemOpen
  \bibfield  {author} {\bibinfo {author} {\bibfnamefont {D.}~\bibnamefont
  {Wecker}}, \bibinfo {author} {\bibfnamefont {M.~B.}\ \bibnamefont
  {Hastings}}, \ and\ \bibinfo {author} {\bibfnamefont {M.}~\bibnamefont
  {Troyer}},\ }\href {\doibase 10.1103/PhysRevA.92.042303} {\bibfield
  {journal} {\bibinfo  {journal} {Phys. Rev. A}\ }\textbf {\bibinfo {volume}
  {92}},\ \bibinfo {pages} {042303} (\bibinfo {year}
  {2015}{\natexlab{b}})}\BibitemShut {NoStop}%
\bibitem [{\citenamefont {Reiner}\ \emph {et~al.}(2019)\citenamefont {Reiner},
  \citenamefont {Wilhelm-Mauch}, \citenamefont {Schön},\ and\ \citenamefont
  {Marthaler}}]{2019}%
  \BibitemOpen
  \bibfield  {author} {\bibinfo {author} {\bibfnamefont {J.-M.}\ \bibnamefont
  {Reiner}}, \bibinfo {author} {\bibfnamefont {F.}~\bibnamefont
  {Wilhelm-Mauch}}, \bibinfo {author} {\bibfnamefont {G.}~\bibnamefont
  {Schön}}, \ and\ \bibinfo {author} {\bibfnamefont {M.}~\bibnamefont
  {Marthaler}},\ }\href {\doibase 10.1088/2058-9565/ab1e85} {\ \textbf
  {\bibinfo {volume} {4}},\ \bibinfo {pages} {035005} (\bibinfo {year}
  {2019})}\BibitemShut {NoStop}%
\bibitem [{\citenamefont {Barison}\ \emph {et~al.}(2021)\citenamefont
  {Barison}, \citenamefont {Vicentini},\ and\ \citenamefont {Carleo}}]{2021}%
  \BibitemOpen
  \bibfield  {author} {\bibinfo {author} {\bibfnamefont {S.}~\bibnamefont
  {Barison}}, \bibinfo {author} {\bibfnamefont {F.}~\bibnamefont {Vicentini}},
  \ and\ \bibinfo {author} {\bibfnamefont {G.}~\bibnamefont {Carleo}},\ }\href
  {\doibase 10.22331/q-2021-07-28-512} {\bibfield  {journal} {\bibinfo
  {journal} {Quantum}\ }\textbf {\bibinfo {volume} {5}},\ \bibinfo {pages}
  {512} (\bibinfo {year} {2021})}\BibitemShut {NoStop}%
\bibitem [{\citenamefont {Hackl}\ \emph {et~al.}(2020)\citenamefont {Hackl},
  \citenamefont {Guaita}, \citenamefont {Shi}, \citenamefont {Haegeman},
  \citenamefont {Demler},\ and\ \citenamefont
  {Cirac}}]{10.21468/SciPostPhys.9.4.048}%
  \BibitemOpen
  \bibfield  {author} {\bibinfo {author} {\bibfnamefont {L.}~\bibnamefont
  {Hackl}}, \bibinfo {author} {\bibfnamefont {T.}~\bibnamefont {Guaita}},
  \bibinfo {author} {\bibfnamefont {T.}~\bibnamefont {Shi}}, \bibinfo {author}
  {\bibfnamefont {J.}~\bibnamefont {Haegeman}}, \bibinfo {author}
  {\bibfnamefont {E.}~\bibnamefont {Demler}}, \ and\ \bibinfo {author}
  {\bibfnamefont {J.~I.}\ \bibnamefont {Cirac}},\ }\href {\doibase
  10.21468/SciPostPhys.9.4.048} {\bibfield  {journal} {\bibinfo  {journal}
  {SciPost Phys.}\ }\textbf {\bibinfo {volume} {9}},\ \bibinfo {pages} {48}
  (\bibinfo {year} {2020})}\BibitemShut {NoStop}%
\bibitem [{\citenamefont {Kolodrubetz}\ \emph {et~al.}(2017)\citenamefont
  {Kolodrubetz}, \citenamefont {Sels}, \citenamefont {Mehta},\ and\
  \citenamefont {Polkovnikov}}]{KOLODRUBETZ20171}%
  \BibitemOpen
  \bibfield  {author} {\bibinfo {author} {\bibfnamefont {M.}~\bibnamefont
  {Kolodrubetz}}, \bibinfo {author} {\bibfnamefont {D.}~\bibnamefont {Sels}},
  \bibinfo {author} {\bibfnamefont {P.}~\bibnamefont {Mehta}}, \ and\ \bibinfo
  {author} {\bibfnamefont {A.}~\bibnamefont {Polkovnikov}},\ }\href {\doibase
  https://doi.org/10.1016/j.physrep.2017.07.001} {\bibfield  {journal}
  {\bibinfo  {journal} {Physics Reports}\ }\textbf {\bibinfo {volume} {697}},\
  \bibinfo {pages} {1} (\bibinfo {year} {2017})},\ \bibinfo {note} {geometry
  and non-adiabatic response in quantum and classical systems}\BibitemShut
  {NoStop}%
\bibitem [{\citenamefont {Bukov}\ \emph {et~al.}(2019)\citenamefont {Bukov},
  \citenamefont {Sels},\ and\ \citenamefont {Polkovnikov}}]{PhysRevX.9.011034}%
  \BibitemOpen
  \bibfield  {author} {\bibinfo {author} {\bibfnamefont {M.}~\bibnamefont
  {Bukov}}, \bibinfo {author} {\bibfnamefont {D.}~\bibnamefont {Sels}}, \ and\
  \bibinfo {author} {\bibfnamefont {A.}~\bibnamefont {Polkovnikov}},\ }\href
  {\doibase 10.1103/PhysRevX.9.011034} {\bibfield  {journal} {\bibinfo
  {journal} {Phys. Rev. X}\ }\textbf {\bibinfo {volume} {9}},\ \bibinfo {pages}
  {011034} (\bibinfo {year} {2019})}\BibitemShut {NoStop}%
\bibitem [{\citenamefont {Du}\ \emph {et~al.}(2020)\citenamefont {Du},
  \citenamefont {Hsieh}, \citenamefont {Liu},\ and\ \citenamefont
  {Tao}}]{PhysRevResearch.2.033125}%
  \BibitemOpen
  \bibfield  {author} {\bibinfo {author} {\bibfnamefont {Y.}~\bibnamefont
  {Du}}, \bibinfo {author} {\bibfnamefont {M.-H.}\ \bibnamefont {Hsieh}},
  \bibinfo {author} {\bibfnamefont {T.}~\bibnamefont {Liu}}, \ and\ \bibinfo
  {author} {\bibfnamefont {D.}~\bibnamefont {Tao}},\ }\href {\doibase
  10.1103/PhysRevResearch.2.033125} {\bibfield  {journal} {\bibinfo  {journal}
  {Phys. Rev. Research}\ }\textbf {\bibinfo {volume} {2}},\ \bibinfo {pages}
  {033125} (\bibinfo {year} {2020})}\BibitemShut {NoStop}%
\bibitem [{\citenamefont {Tacchino}\ \emph {et~al.}(2020)\citenamefont
  {Tacchino}, \citenamefont {Chiesa}, \citenamefont {Carretta},\ and\
  \citenamefont {Gerace}}]{https://doi.org/10.1002/qute.201900052}%
  \BibitemOpen
  \bibfield  {author} {\bibinfo {author} {\bibfnamefont {F.}~\bibnamefont
  {Tacchino}}, \bibinfo {author} {\bibfnamefont {A.}~\bibnamefont {Chiesa}},
  \bibinfo {author} {\bibfnamefont {S.}~\bibnamefont {Carretta}}, \ and\
  \bibinfo {author} {\bibfnamefont {D.}~\bibnamefont {Gerace}},\ }\href
  {\doibase https://doi.org/10.1002/qute.201900052} {\bibfield  {journal}
  {\bibinfo  {journal} {Advanced Quantum Technologies}\ }\textbf {\bibinfo
  {volume} {3}},\ \bibinfo {pages} {1900052} (\bibinfo {year}
  {2020})}\BibitemShut {NoStop}%
\bibitem [{\citenamefont {ANIS}\ \emph {et~al.}(2021)\citenamefont {ANIS},
  \citenamefont {Abraham}, \citenamefont {AduOffei}, \citenamefont {Agarwal},
  \citenamefont {Agliardi},\ and\ \citenamefont {et. al.}}]{Qiskit}%
  \BibitemOpen
  \bibfield  {author} {\bibinfo {author} {\bibfnamefont {M.~S.}\ \bibnamefont
  {ANIS}}, \bibinfo {author} {\bibfnamefont {H.}~\bibnamefont {Abraham}},
  \bibinfo {author} {\bibnamefont {AduOffei}}, \bibinfo {author} {\bibfnamefont
  {R.}~\bibnamefont {Agarwal}}, \bibinfo {author} {\bibfnamefont
  {G.}~\bibnamefont {Agliardi}}, \ and\ \bibinfo {author} {\bibfnamefont
  {M.~A.}\ \bibnamefont {et. al.}},\ }\href {\doibase 10.5281/zenodo.2573505}
  {\enquote {\bibinfo {title} {Qiskit: An open-source framework for quantum
  computing},}\ } (\bibinfo {year} {2021})\BibitemShut {NoStop}%
\bibitem [{\citenamefont {Bravyi}\ and\ \citenamefont
  {Kitaev}(2002)}]{BRAVYI2002210}%
  \BibitemOpen
  \bibfield  {author} {\bibinfo {author} {\bibfnamefont {S.~B.}\ \bibnamefont
  {Bravyi}}\ and\ \bibinfo {author} {\bibfnamefont {A.~Y.}\ \bibnamefont
  {Kitaev}},\ }\href@noop {} {\bibfield  {journal} {\bibinfo  {journal} {Annals
  of Physics}\ }\textbf {\bibinfo {volume} {298}},\ \bibinfo {pages} {210}
  (\bibinfo {year} {2002})}\BibitemShut {NoStop}%
\bibitem [{\citenamefont {Tranter}\ \emph {et~al.}(2018)\citenamefont
  {Tranter}, \citenamefont {Love}, \citenamefont {Mintert},\ and\ \citenamefont
  {Coveney}}]{tranter2018}%
  \BibitemOpen
  \bibfield  {author} {\bibinfo {author} {\bibfnamefont {A.}~\bibnamefont
  {Tranter}}, \bibinfo {author} {\bibfnamefont {P.~J.}\ \bibnamefont {Love}},
  \bibinfo {author} {\bibfnamefont {F.}~\bibnamefont {Mintert}}, \ and\
  \bibinfo {author} {\bibfnamefont {P.~V.}\ \bibnamefont {Coveney}},\
  }\href@noop {} {\bibfield  {journal} {\bibinfo  {journal} {Journal of
  Chemical Theory and Computation}\ }\textbf {\bibinfo {volume} {14}},\
  \bibinfo {pages} {5617} (\bibinfo {year} {2018})}\BibitemShut {NoStop}%
\bibitem [{Note1()}]{Note1}%
  \BibitemOpen
  \bibinfo {note} {See \protect \url
  {https://quantum-computing.ibm.com/.}}\BibitemShut {Stop}%
\bibitem [{\citenamefont {Nation}\ \emph {et~al.}(2021)\citenamefont {Nation},
  \citenamefont {Kang}, \citenamefont {Sundaresan},\ and\ \citenamefont
  {Gambetta}}]{PRXQuantum.2.040326}%
  \BibitemOpen
  \bibfield  {author} {\bibinfo {author} {\bibfnamefont {P.~D.}\ \bibnamefont
  {Nation}}, \bibinfo {author} {\bibfnamefont {H.}~\bibnamefont {Kang}},
  \bibinfo {author} {\bibfnamefont {N.}~\bibnamefont {Sundaresan}}, \ and\
  \bibinfo {author} {\bibfnamefont {J.~M.}\ \bibnamefont {Gambetta}},\ }\href
  {\doibase 10.1103/PRXQuantum.2.040326} {\bibfield  {journal} {\bibinfo
  {journal} {PRX Quantum}\ }\textbf {\bibinfo {volume} {2}},\ \bibinfo {pages}
  {040326} (\bibinfo {year} {2021})}\BibitemShut {NoStop}%
\bibitem [{\citenamefont {Neill}\ \emph {et~al.}(2021)\citenamefont {Neill}
  \emph {et~al.}}]{neill_accurately_2021}%
  \BibitemOpen
  \bibfield  {author} {\bibinfo {author} {\bibfnamefont {C.}~\bibnamefont
  {Neill}} \emph {et~al.},\ }\href@noop {} {\bibfield  {journal} {\bibinfo
  {journal} {Nature}\ }\textbf {\bibinfo {volume} {594}},\ \bibinfo {pages}
  {508} (\bibinfo {year} {2021})}\BibitemShut {NoStop}%
\bibitem [{\citenamefont {Fetter}\ and\ \citenamefont
  {Walecka}(1971)}]{Fetter}%
  \BibitemOpen
  \bibfield  {author} {\bibinfo {author} {\bibfnamefont {A.~L.}\ \bibnamefont
  {Fetter}}\ and\ \bibinfo {author} {\bibfnamefont {J.~D.}\ \bibnamefont
  {Walecka}},\ }\href@noop {} {\emph {\bibinfo {title} {Quantum Theory of
  Many-Particle Systems}}}\ (\bibinfo  {publisher} {McGraw-Hill},\ \bibinfo
  {address} {Boston},\ \bibinfo {year} {1971})\BibitemShut {NoStop}%
\end{thebibliography}%


\end{document}